\documentclass[journal]{IEEEtran}

\usepackage[utf8]{inputenc}
\usepackage{mathtools}
\usepackage{dirtytalk}
\usepackage{caption}
\usepackage{subcaption}

\usepackage{xcolor}
\mathchardef\mhyphen="2D

\newcommand{\name}{LINTS$^{\rm{RT}}$} % LaPoint$^{RT}

\begin{document}
\title{\name: A Learning-driven Testbed for Intelligent Scheduling in Embedded Systems}
\author{
\IEEEauthorblockN{Zelun~Kong,
        Yaswanth~Yadlapalli,~
        Soroush~Bateni,~
        Junfeng~Guo,~
        Cong~Liu} \\
\IEEEauthorblockA{\textit{Department of Computer Science,} \\
\textit{The University of Texas at Dallas}\\}}
\maketitle
\thispagestyle{plain}
\pagestyle{plain}
\begin{abstract}
    Due to the increasing complexity seen in both workloads and hardware resources in state-of-the-art embedded systems, developing efficient real-time schedulers and the corresponding schedulability tests becomes rather challenging. Although close to optimal schedulability performance can be achieved for supporting simple system models in practice, adding any small complexity element into the problem context such as non-preemption or resource heterogeneity would cause significant pessimism, which may not be eliminated by any existing scheduling technique. In this paper, we present \name{}, a learning-based testbed for intelligent real-time scheduling, which has the potential to handle various complexities seen in practice. The design of \name{} is fundamentally motivated by AlphaGo Zero for playing the board game Go, and specifically addresses several critical challenges due to the real-time scheduling context. We first present a clean design of \name{} for supporting the basic case: scheduling sporadic workloads on a homogeneous multiprocessor, and then demonstrate how to easily extend the framework to handle further complexities such as non-preemption and resource heterogeneity. Both application and OS-level implementation and evaluation demonstrate that \name{} is able to achieve significantly higher runtime schedulability under different settings compared to perhaps the most commonly applied schedulers, global EDF, and RM. To our knowledge, this work is the first attempt to design and implement an extensible learning-based testbed for autonomously making real-time scheduling decisions.
\end{abstract}

% \begin{IEEEkeywords}

% \end{IEEEkeywords}

% \fi
\IEEEpeerreviewmaketitle

\section{INTRODUCTION}
\label{sec:intro}

%\Cong{Potential improvements: 1. Regarding guarantee: at least providing average-case prediction using DNN. What else we can do? 2. }

One of the core areas within the embedded and real-time systems research community is to design efficient and practical rule-based scheduling algorithms (e.g., EDF) for workloads characterized by various formal task models. For earlier task and system models, rather elegant scheduling algorithms along with their corresponding schedulability tests have been proposed, e.g., the optimality of EDF on a uniprocessor~\cite{liu2000}. Unfortunately, it becomes increasingly hard to design such scheduling tools due to the increasing complexities seen in both workloads and hardware resources in state-of-the-art embedded systems. Any tiny piece of complexity in either workload (e.g., non-preemption or data dependency) or hardware (e.g., resource heterogeneity) would cause significant pessimism in using existing scheduling disciplines. For instance, the non-preemptive scheduling of sporadic task systems on a homogeneous multiprocessor has shown to be hard~\cite{baruah2006non,  baek2018non, lee2017improved}, where using existing scheduling algorithms may result in unacceptable utilization loss under various scenarios. Adding multiple complexity elements into the real-time scheduling context would make most existing rule-based scheduler design fail. For instance, the problem of non-preemptively scheduling sporadic tasks on a heterogeneous multiprocessor largely remains as an open problem, despite only two straightforward complexity elements (i.e., non-preemption and resource heterogeneity) being considered.

This work is motivated due to a collaboration with a prominent industrial partner in the field of wireless baseband networking systems. We were asked a challenging question that is it possible to design a real-time scheduling framework that would be efficient in terms of achieving low deadline miss ratio\footnote{In many industrial systems including wireless baseband, it is often not possible and thus not required to guarantee stringent hard real-time correct, i.e., every single deadline must be met. Many such systems allow deadline misses, however, such deadline misses shall be rather infrequent (e.g., $<10\%$). Note that different application domains have different performance criteria.} in handling workloads that may exhibit various complexity characteristics.
Unfortunately, we realize that it is difficult (if not impossible) to design such a scheduling solution containing a set of fixed rules or heuristics that may cover various system settings exhibiting different workload and hardware characteristics.  
Indeed, many of these real-time scheduling problems are NP-hard problems, and to come up with a general base framework using carefully chosen rule-based approaches for these problems may not be feasible. An alternative to solve traditional NP-hard problems is by using machine learning. In particular, reinforcement learning (RL) has been applied for many such problems, with a recent famous example solution on playing the Go board game. The state-of-the-art RL-driven solution is called AlphaGo Zero~\cite{silver2017mastering,silver2016mastering}, which uses a Monte Carlo tree search to find moves based on previously “learned” moves. 

Driven by the issues raised above, we present \name, a learning-based testbed for intelligent real-time scheduling. \name contains three major components: Simulator, Deep Neural Network (DNN), and Monte Carlo Tree Search (MCTS). During the training phase given any training task set, the simulator simulates a sporadic multiprocessor system. By looking at the current state of the simulator, the DNN determines the probabilities of future actions. Then MCTS randomly selects a finite set of actions guided by the action probabilities from the DNN. The simulator then simulates by applying the actions from MCTS. This procedure is repeated until we reach an end state upon which a reward is calculated. This reward is sent as feedback to the DNN. DNN associates the reward with all the intermediate states and actions between the start and end states. This process is then repeated for all the task sets in the training data. The trained DNN from this procedure is used as a dynamic priority scheduler in the decision phase. Hence, the structural design and implementation of the DNN have to be taken into consideration due to the tradeoff between expressibility vs. overhead as running a DNN to make runtime decisions could cause an unacceptable latency penalty. This three-component system can be easily extended to take into consideration additional constraints, as shown in Sec.~\ref{sec:extensions}. 

We have implemented \name{} both as an application-level simulator and a scheduler plugin within a real-time OS, i.e., LITMUS$^{\rm{RT}}$~\cite{brandenburg2011scheduling, calandrino2006litmus}. 
Through extensive evaluation, \name{} is proven to have the following properties:

\begin{itemize}
    \item \textbf{Efficacy.} We compare the schedulability of various task sets by \name in simulated environments considering practical scheduling-induced overheads as well as in LITMUS$^{RT}$. We observe significant improvements w.r.t. runtime schedulability under both settings in comparison to GEDF and RM, for scheduling real-time sporadic task sets considering complexities expressed via non-preemption and resource heterogeneity. 
    \item \textbf{Extensibility.} We demonstrate how to easily extend the basic framework of \name, which supports sporadic tasks scheduled on a homogeneous multiprocessor in a preemptive manner, to incorporate non-preemption and/or resource heterogeneity, which represents a practical yet difficult characteristic of workloads and hardware, respectively. Through such demonstration on extensibility, we aim to encourage other researchers from both academia and industry to leverage and extend \name{} in addressing their specific scheduling problems.
    \item \textbf{Practicality.} Our implementation of \name{} is shown to be practical. Due to an optimized structure and GPU-based implementation of the DNN component, \name{} incurs rather low latency overhead when making runtime scheduling decisions.
\end{itemize}

To our knowledge, this work represents the first attempt to design and implement an extensible learning-based framework for autonomously making real-time scheduling decisions. A few limitations of our focused settings are worth noting. First, while we believe that the design of \name{} can apply to other scheduling disciplines, \name{} focuses on global scheduling, which is shown to be superior under many common settings~\cite{bastoni2010empirical, brandenburg2011scheduling}. In future work, we will extend \name{} to consider other disciplines. Similarly, in creating \name{}, producing a fully-featured testbed covering all complexities seen in practice would simply not be feasible at present. Rather, our goal was to produce an easily extensible platform which could be further leveraged and extended by other researchers. For this reason, our design and implementation have focused on independent sporadic tasks. We leave issues such as support for I/O, shared resources, and other workload characteristics  such as dependency and parallelism as future work.

\vspace{-2mm}
\section{BACKGROUND}
\label{sec:background}

\subsection{Real-time Scheduling}
A rich set of real-time scheduling algorithms and their corresponding schedulability tests have been proposed in the literature, including a large body of such works focusing on scheduling sporadic real-time tasks on a homogeneous multiprocessor. 
Till recently, this classical scheduling problem is deemed to be practically resolved as a heuristic-based algorithm is shown to be able to achieve a rather high runtime schedulability considering real system overheads~\cite{brandenburg2016global}. 
Besides this basic system model (i.e., sporadic workloads preemptively scheduled on homogeneous multiprocessor), extensive efforts have been made to handle increasing complexities seen in the task model, such as non-preemptive execution~\cite{baruah2006non, baek2018non, lee2017improved},  precedence constraints~\cite{saifullah2014parallel, jiang2017semi, jiang2019decomposition, ueter2018reservation}, self-suspensions~\cite{chen2019many, dong2016closing}, and mixed-criticality~\cite{li2012outstanding, baruah2014mixed, guo2017sustainability, papadopoulos2018adaptmc}, as well as those seen in the hardware platform, such as heterogeneous multiprocessors~\cite{baruah2015real, singh2016real}, computing accelerators~\cite{elliott2015real, biondi2016framework, capodieci2018deadline}, and shared resources~\cite{biondi2015schedulability, huang2016resource, biondi2016lightweight}. Unfortunately, although promising progress has been made for dealing with various complexities, many such complexities yield rather pessimistic results. For instance, the non-preemptive scheduling of sporadic task systems on a homogeneous multiprocessor, which merely adds the complexity of non-preemptive execution into the scheduling context, remains largely as an open problem since the state-of-the-art schedulability tests~\cite{baruah2006non,  baek2018non, lee2017improved} may cause significant utilization loss under many settings. When facing a system environment where  multiple such complexities are presented, the real-time scheduling problem may become significantly difficult. For instance, no efficient scheduling technique is known to be  able to non-preemptively schedule just simple sporadic workloads on a heterogeneous multiprocessor where cores have different speeds.

This observed difficulty in designing efficient rule-based real-time scheduling algorithms motivates us to develop \name. 
\name{} focuses on global scheduling, which has been shown to be effective and superior under several common settings compared to other disciplines such as partitioned and clustered scheduling~\cite{davis2011survey}.

\subsection{Machine Learning Techniques}

\begin{figure}[t]
    \centering
    \includegraphics[width=0.8\columnwidth]{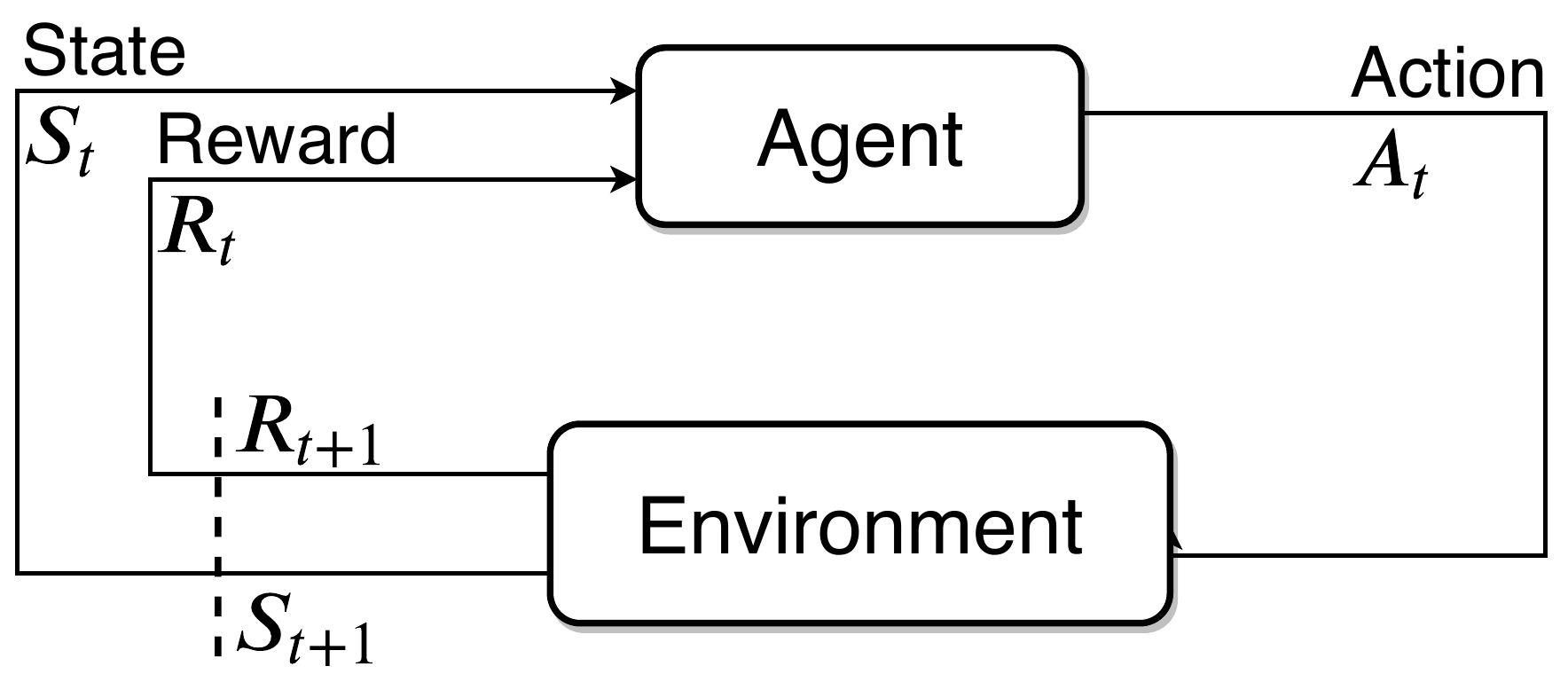}
    \caption{Reinforcement learning.}
    \vspace{-3mm}
    \label{fig:reinforcement_learning}
\end{figure}

% A typical machine learning program contains 2 phases, a training phase where the program tries to learn the structure of the problem and comes up with a possible hypothesis, and the decision phase were given an input the program gives out the output based on the learned hypothesis.

Reinforcement learning (RL) is a branch of machine learning, which is particularly good at training an intelligent agent that can act autonomously in a specific environment. The trained autonomous agent is supposed to take proper actions through interaction with the environment and improve the “goodness” of the system state over time. Figure~\ref{fig:reinforcement_learning} illustrates the structure of a reinforcement learning system. Reinforcement learning systems are composed of four elements:

\begin{itemize}
    \item State. For example, the system state of our scheduling problem, denoted as $S$. In our scheduling problem, the job queue and processor information (heterogeneity) constitute $S$.
    \item Actions. Given the current state, only a subset of all actions is allowed. For example, in our scheduling problem, only jobs that are already in the job waiting list and job running list are allowed to be selected and run in the next time slot. Actions are denoted as $a$. 
    \item Reward $r$. After the agent takes some actions, the reinforcement learning system needs to interact with the environment and evaluate the utility of the action. The utility values will be used to generate a reward to help the reinforcement learning system to train the intelligence agent. For example, scheduling actions that make some jobs miss their deadline will have a low reward, whereas scheduling actions that achieve the opposite will have a higher reward. 
    \item Policy. A mapping from states to actions denoted as $\pi: S \rightarrow A$. It is the only output of the reinforcement learning system. Note that a policy could be a table, a function, or even a neural network.
\end{itemize}

Detailed state, action, and reward definitions for our system are given in Section~\ref{sec:SARdef}. 
RL enables the system to keep continuously tweaking its learned hypothesis as the system progresses through its computation. For our scheduling system, we define the state to be the current job queue and actions as selecting a job for execution. But, setting an immediate reward for every action the scheduler takes is challenging because it is not immediately apparent in the system how the granular scheduling decisions affect the overall schedule. Coincidentally learning systems for board games such as Go also run into a similar problem, where the reward for every move is unclear, as it is not immediately apparent in the game how a move affects the outcome of the game. \textit{AlphaGo Zero}, which is a state-of-the-art RL algorithm for playing Go, inspires our approach.

\section{DESIGN of \name}
\label{sec:design}

\begin{figure}[t]
    \centering
    \includegraphics[width=\columnwidth]{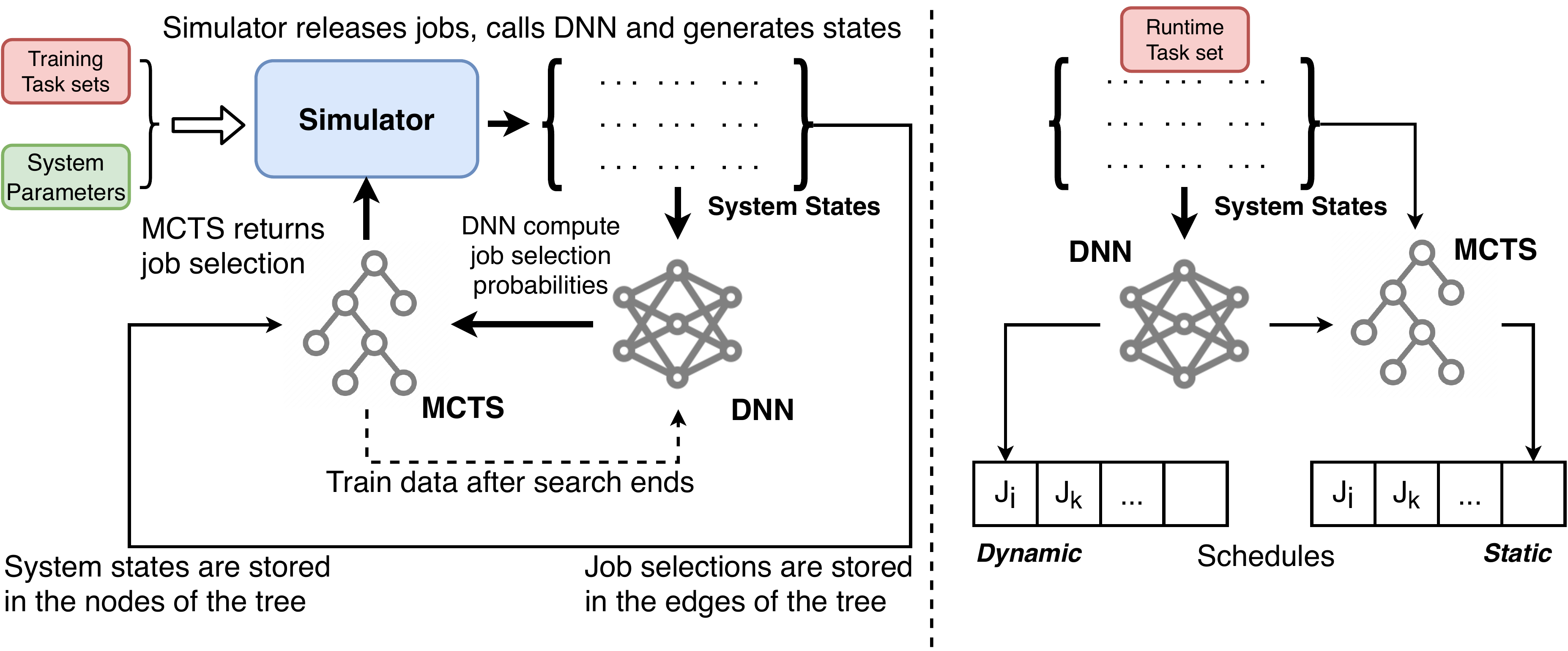}
    \caption{Overview of \name{}, with the training phase and runtime scheduling phase illustrated on the left and right subfigure, respectively.}
    \label{fig:overview}
\end{figure}

\name{} utilizes an optimization search algorithm which originates from \textit{AlphaGo Zero}. However, there is a set of inherent differences between the environment of real-time scheduling and the board game Go, which makes it non-trivial to leverage the fundamental reinforcement learning idea.
% \begin{itemize}
    % \item 
    \noindent\textbf{Latency sensitivity.} A major difference is regarding latency sensitivity, as runtime actions on playing the board game Go do not need to be real-time, whereas real-time scheduling decisions have to be made in a stringent real-time fashion. This causes a key challenge for leveraging any reinforcement learning technique (particularly those using deep neural networks) in the scheduling problem context because such a learning-based decision-making process often incurs a non-trivial latency overhead, unlike fast heuristic-base schedulers such as EDF and RM. Since the game Go does not have such strict latency constraints, the DNN deployed in AlphaGo Zero utilizes ResNet~\cite{he2016deep}, which may incur a runtime overhead of at least 100ms~\cite{he2016deep}. On the other hand, using a small but fast DNN may result in unsatisfactory accuracy, which may cause frequent deadline misses. We carefully explore the tradeoff between accuracy and latency when designing and implement \name. As discussed in Sec.~\ref{sec:DNNTrain} in detail, we develop a new DNN structure, which yields high accuracy while incurring a rather small runtime overhead, as reported in the evaluation. 
    % \item 
    
    \noindent\textbf{Adversary.} Board games are played against an opponent, where one could improve the performance of the algorithm by self-play. It is thus straightforward to define a reward function as the win condition is clear. In contrast, there is no such adversary for real-time scheduling problems, which makes it challenging to define and compute rewards as there isn't any intuitive win condition for sporadic task scheduling. To resolve this issue, we define the reward for an action as the longest possible time for the sporadic task system not to miss a deadline after taking the action. To compute this reward, we implement an efficient scheduling simulator, which we describe in Sec.~\ref{sec:mcts}.    
    % \item 
    
    \noindent\textbf{Action predictability.} Predicting moves of a deterministic board game such as Go is vastly simpler than predicting a real-time system because the system has hidden variables such as the unpredictable job releases of sporadic tasks that results in non-deterministic behavior. To leverage the reinforcement learning framework for the sporadic real-time task scheduling problem, it is thus required to obtain more comprehensive and reliable training data. Only well-distributed training data can ensure that the neural network would learn the pattern of the sporadic tasks. This is because the job release time and execution time of sporadic tasks are usually subject to random distributions. Therefore, if the training data is not comprehensive enough, the distribution cannot be learned properly, leading to improper scheduling decisions. This is illustrated in Fig.~\ref{fig:policy_iteration} and discussed in detail in Sec.~\ref{sec:DNNTrain}.
    % \item 
    
    \noindent\textbf{Operation duration.} Real-time scheduling, especially for sporadic tasks, is infinite in length, whereas any board game is finite. This infinity property could lead to error accumulation, i.e., a former improper scheduling decision is very likely to cause a series of deadline misses at later times, which is a very hard problem. A perfect solution to resolve this challenge is to guarantee that optimal scheduling decisions are made at every instant, which is clearly impossible. Our alternative solution is to design a DNN to output an evaluation score that measures the \say{optimality} of the current state. \name{} will then make scheduling decisions that can increase this score accordingly. Details are discussed in Sec.~\ref{sec:decision_static} and Sec.~\ref{sec:decision_dynamic}.
% \end{itemize}
These differences make it challenging from many aspects to developing a learning-based framework for making real-time scheduling decisions. To overcome these challenges, the design and implementation of \name{} have been carefully optimized and tailored specifically to the problem context of real-time scheduling.
\subsection{Overview}
The overview of \name{} is illustrated in Fig \ref{fig:overview}, which has been generally described in Sec.~\ref{sec:intro}.  

For the training phase, the input for \name{} is a set of training task sets, and the output would be a State-Action mapping.
A task set from the input is given to the simulator, which then simulates the job releases and constructs the state as a matrix (also called tensor) derived from the current job queue and processor information. This state contains all the information that a rule-based scheduler would require to make a scheduling decision. The Action is defined as a tensor that represents a job being selected for execution.

The reward for an action is defined as the maximum runtime achievable without missing a deadline. Through our design, this reward can be estimated by examining multiple executions of the Monte Carlo Tree Search simulation and finding scenarios that include that specific action.

However, the inherently stochastic nature of the MCTS makes it very inefficient. To remedy this problem, we use a Deep Neural Network as a companion to the MCTS in order to reduce the randomness of the tree search. The DNN would be trained to take the state tensor as an input and give a reasonable action as the output. This training should be possible because the state already contains all the necessary information to make a scheduling decision (e.g., a rule-based scheduler can make decisions based on the same state).

As we shall discuss in Sec.~\ref{sec:mcts}, we use the DNN as a guide to the MCTS to limit the search to known \say{good} areas of the search space. This search is continued until the MCTS finds a feasible schedule (or an iteration threshold is reached) for a task set. The output of this search would in-turn, be used to train the DNN. This process is repeated for all the task sets in the input.

Both the DNN guided MCTS simulations and the DNN by itself are different versions of State-Action mapping. This means that a well-trained DNN can potentially be used alone to make runtime scheduling decisions. However, the simulation-based DNN guided MCTS is much more accurate, as we shall discuss in Sec.~\ref{sec:DNNTrain}, albeit with a larger overhead. To explore this tradeoff, we design \name{} under two configurations: a dynamic configuration, suitable for dynamic workloads that only uses the DNN, and a static configuration which uses the entirety of the DNN-guided MCTS, and is thus more suitable for static scenarios where workload parameters are pre-known, thus allowing \name{} to generate a scheduling table offline and make runtime scheduling decisions accordingly. We have evaluated both configurations in the evaluation.

\subsection{System Model}
% System
We consider the problem of scheduling a set of $n$ sporadic tasks $\tau = \{\tau_1, \tau_2, \cdots, \tau_n\}$ on $m$ identical processors (homogeneous setting) or processors with varying speeds (heterogeneous setting) in a preemptive or non-preemptive manner. 
As discussed before, our current design scope focuses on global scheduling, where the jobs running on a processor can be migrated to another processor by the scheduler with certain overhead. 
% Task
Each task $\tau_i$ is specified by a vector $\vec{\tau_i} = (p_i, d_i, e_i)$, where $p_i$ denotes the minimum interval between two jobs of a task, $e_i$ denotes the worst-case execution time for a job, and $d_i$ is the relative deadline ($d_i = p_i$).
% Job
Let $J_i^{(j)}$ denote the $j$-th instance of task $\tau_i$, which can also be described by a vector $\vec{J_i} ^{\small{{(j)}}} = (R_i, E_i, D_i)$, where $R_i$ is the release time, $E_i$ is the actual execution time, and $D_i$ is the absolute deadline. 

% time slot
A time slot is considered to be the basic scheduling time unit. A job must execute during a whole time slot or not execute in that time slot at all. 
% Job lists
A job has two different states: waiting and running. Jobs with the same status are stored respectively in two different job lists, \textit{job waiting list} $L_w$ and \textit{job running list} $L_r$. Let $N_l = (\|L_r\| + \|L_w\|)$ where $\|L_r\|,\|L_w\|$ are the lengths of $L_w$ and $L_r$ respectively.
Note that these lists are disjoint as one job can only be in one of the lists.

\subsection{State, Action and Reward Representation}
\label{sec:SARdef}

\subsubsection{State} 
At the start of the time slot $[t, t+1)$ with $i$ idle processors is represented as $\vec{s}_{t, i}$. The current status of the system can be represented as a tensor of size $N_l \times 3$, in which every job in $L_r$ and $L_w$ is represented by a row. The first column of this tensor is the relative deadline of the job, the second column is the remaining execution time of the job, and the third column is a boolean which is true if the job is executing in time slot $[t, t+1)$. For example, the following can be a valid representation of the system's status:
$$
    \begin{bmatrix}
        D_{i_1}^{(j_1)} - t & E_{i_1}^{(j_1)} - \mathrm{execution}(R_{i_1}^{(j_1)}, t) & 0/1  \\
        D_{i_2}^{(j_2)} - t & E_{i_2}^{(j_2)} - \mathrm{execution}(R_{i_2}^{(j_2)}, t) & 0/1  \\
        \vdots & \vdots \\
        D_{i_n}^{(j_n)} - t & E_{i_n}^{(j_n)} - \mathrm{execution}(R_{i_n}^{(j_n)}, t) & 0/1
    \end{bmatrix}
    \label{eq:preemptive_system_state}
$$
This tensor can be calculated from $L_r$ and $L_w$. Additionally, we keep the last $N_h$ statuses of the current tasks in the state (this improves the performance of the DNN while making scheduling decisions described in section \ref{sec:DNNTrain}). Thus, we construct a state tensor at time point $t$ of size $N_l \times 3 \times N_h$.

\subsubsection{Action} 
Action is defined as a boolean vector of length $N_l$, where only one entry can be true. 
\begin{equation}
    \label{eq:action}
    \vec{a}_{t, i} = (0, \cdots, 1, \cdots, 0)
\end{equation}
Here $t$ and $i$ have the same meaning as they are in the system state $\vec{s}_{t, i}$. 
One action corresponds to choosing one job from either $L_w$ or $L_r$ to execute in the next time-slot. Therefore, for an $m$ processor system, given the current state, we need to take at most $m$ actions with $m-1$ intermediate states (i.e., we use $\{(\vec{s}_{t, m}, \vec{a}_{t, m}), (\vec{s}_{t, m - 1}, \vec{a}_{t, m - 1}), \cdots, (\vec{s}_{t, 1}, \vec{a}_{t, 1})\}$ to represent the jobs running in the next time slot). These intermediary states are constructed by removing the jobs that have been selected by the previous action.

\subsubsection{Reward}
\label{sec:reward}
The reward value computation method is inspired by AlphaGo Zero, which uses MCTS. MCTS is a simulation-based algorithm which uses a tree where each node represents a state, and each edge represents a possible action to take. Also, a \textit{policy} $p$ is defined as the probability distribution over the set of actions given a state $= p(a | s)$. Additionally, each node in the tree is assigned \textit{value} $v$ to represent the goodness of a state. Initially $p$ and $v$ are completely uniform. However, these distributions are updated by performing \textit{roll-outs}. A roll-out is defined as simulating random actions (sampled from the probability distribution defined by $p$ and $v$) from the current node until the end condition of the system, which for us is either a deadline miss or when the time elapsed is greater than the hyper-period of the task set.

For example, a roll-out from the initial state (root of the tree) simulates the schedule from $\vec{s}_{0, m} \rightarrow \vec{s}_{t, m}$ where either $t > H(\tau)$ or at $t$, one of the jobs misses its deadline. The outcome of a roll-out contributes to $v$ and $p$ of nodes and edges respectively in the path of the roll-out. These updates make the subsequent roll-outs explore the \say{good} areas of the search tree as the random actions are sampled from the probability distributions.

For this approach to work, we need to create a scheduling simulator to perform this roll-out, the simulator implementation is defined in Sec. \ref{sec:imp_details}. We define the reward for an action as the maximum system runtime out of all the roll-outs that go through this action. Note that our reward is always a positive number, and the RL framework is capable of identifying the best scheduling decision using the variation in the magnitude of reward. However, the inherently stochastic nature of MCTS makes it very inefficient. It might take thousands of roll-outs until the algorithm starts to search the \say{good} areas. It would be beneficial if the tree search had a guide to help it find good states to expand.

\subsection{Deep Neural Network guided MCTS}
\label{sec:mcts}
To guide the MCTS algorithm, we need a component which given a state $\vec{s}_{t, i}$ produces a reasonable action to take. Since the state contains all the required information that a rule-based scheduler requires to make a scheduling decision (for sporadic tasks on a homogeneous multiprocessor system), we should be able to train a Deep Neural Network (DNN) to do the same. Let $\theta$ be the parameter of DNN $f$, then the DNN can be written as  $f_\theta(\vec{s}_{t,i}) = (\vec{p}, v) $ where:
\begin{itemize}
    \item $\vec{p}$ is the policy, which represents the probability of a job being selected to run in the next time slot for the given system state $\vec{s}_{t, i}$. 
    For example,
    $$
        \vec{p} = (0.01, 0.03, 0.92, \cdots).
    $$
    \item $v \in [-1,1]$ is evaluation value, which represents the \say{goodness} of state $\vec{s}_{t, i}$. A value of $1$ means that in the foreseeable future, no jobs will miss its deadline. A value of $-1$ means that in the next time slot, some job misses its deadline. 
\end{itemize}
The DNN achieves the same functionality as the State-Action mapping, which is the output from RL. Therefore, we can use the overall RL model to train the DNN over time. For each iteration of the RL model, the DNN would keep getting better at guiding the MCTS. The details about the DNN's structure and training are discussed in the next section. For now, assume that we have a DNN that is semi-competent. Next, we describe how such a DNN can be used to guide the MCTS. 
% (see Fig.~\ref{fig:dnn_guided_mcts}).

In the MCTS tree, each edge $(\vec{s}_{t, i}, \vec{a}_{t, i}
)$ need to store 3 values: prior probability $
P(\vec{s}_{t, i}, \vec{a}_{t, i}) = \vec{p} \boldsymbol{\cdot} \vec{a}_{t, i}$ where $f_{\theta}(\vec{s}_{t,i}) = (\vec{p}, v)$, visit count $N(\vec{s}_{t, i}, \vec{a}_{t, i})$ which is equal to the number of times the edge has been visited in previous roll-outs, and the job selection value $Q(\vec{s}_{t, i}, \vec{a}_{t, i})$ which is initialized to zero. Each roll-out starts from the initial state $\vec{s}_{0, m}$, then iteratively selects the actions that maximize an upper confidence bound $Q(\vec{s}_{t, i}, \vec{a}_{t, i}) + U(\vec{s}_{t, i}, \vec{a}_{t, i})$, where 
\begin{equation}
    U(\vec{s}_{t, i}, \vec{a}_{t, i}) \propto \frac{P(\vec{s}_{t, i}, \vec{a}_{t, i})}{1 + N(\vec{s}_{t, i}, \vec{a}_{t, i})}.
\end{equation}

The simulation stops when it encounters a leaf node $\vec{s}_{t',j}$ (i.e., when a job misses its deadline or $t' > H(\tau)$). For each edge $(\vec{s}_{t, i}, \vec{a}_{t, i})$ traversed in the roll-out, $N(\vec{s}_{t, i}, \vec{a}_{t, i})$ is incremented by 1 and:
\begin{equation}
    Q(\vec{s}_{t, i}, \vec{a}_{t, i}) = \frac{1}{N(\vec{s}_{t, i}, \vec{a}_{t, i})} \sum_{\vec{s} \in \{\vec{s}_{t,i} \rightarrow \vec{s}_{t',j} \} } V(\vec{s})
    \label{eq:value_update}
\end{equation}
where $\{\vec{s}_{t,i} \rightarrow \vec{s}_{t',j} \}$ indicates the set of states in the path from the current node to the leaf node and $V(\vec{s}) = v$ where $f_{\theta} (\vec{s}) = (\vec{p}, v)$. 

\begin{figure}[t]
    \centering
    \includegraphics[width=1\columnwidth]{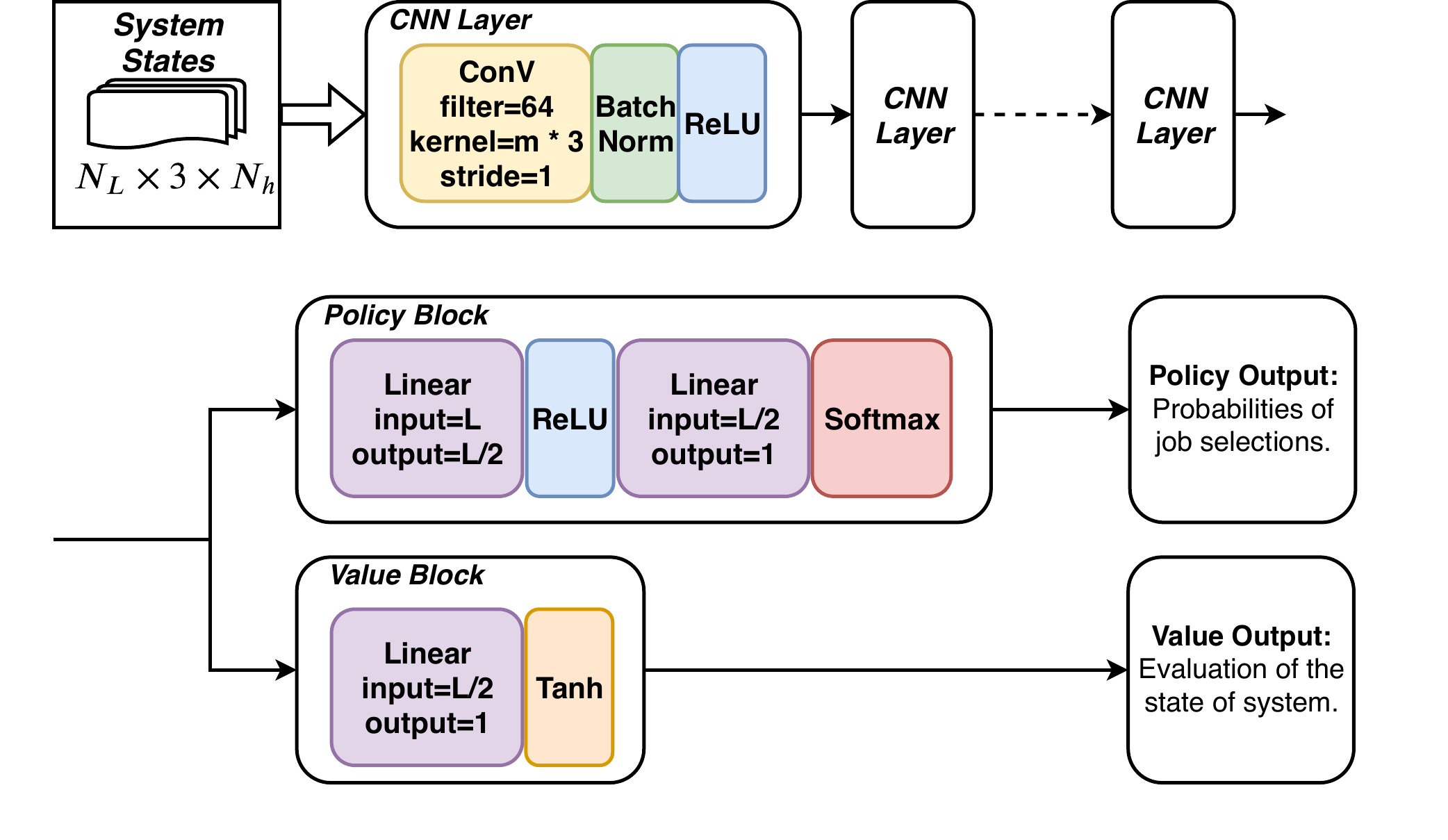}
    \caption{Network architecture. }
    \label{fig:network_architecture}
\end{figure}

\subsection{DNN Structure and Training}
\label{sec:DNNTrain}

\subsubsection{Deep Neural Network Architecture}
Figure~\ref{fig:network_architecture} shows the architecture of the DNN used in \name.
The network is mainly composed of three modules: the input module, the policy block, and the value block. 
The input module is composed of 5 CNN blocks~\cite{krizhevsky2012imagenet}, with each block containing a convolutional layer with 64, $m\times3$, 1 as its filter size, kernel size and stride respectively followed by a batch normalization and ReLU layers. The policy block is composed of 3 linear layers and uses the $softmax$ as the activation function as each element in the policy lies in the range of [0,1].
The value block contains 2 linear layers and uses $Tanh$ as the activation function as the evaluation value $v \in [-1, 1]$. Note that the structure of DNN is purposefully kept lean as compared to DNNs for other applications such as Go, to minimize the scheduling overhead. 

\subsubsection{Policy Iteration and Network Training}
Here, the DNN-guided MCTS may be seen as a trial-and-error algorithm which, given the DNN parameters $\theta$ and a root state $\vec{s}_{t, i}$, computes the State-Action mapping as a probability tensor $\vec{\pi}$. 
Note that $\vec{\pi}$ and $\vec{p}$ are both representing the probability of a job being selected at state $\vec{s}_{t, i}$. However, as $\vec{\pi}$ is computed from simulating many roll-outs, it is more accurate and reliable than $\vec{p}$, especially when the DNN has not completed training.

For a state $\vec{s}_{t,i}$, we can compute $\vec{\pi}$ from the DNN-guided MCTS and then compute $z$ as the number of time slots from $t$ where no job misses its deadline. The $\vec{\pi}, z$ pair can then be used as a training sample to update the parameters $\theta$ of the DNN, making the output job selection probabilities and evaluation value, $f_\theta(\vec{s}_{t, i}) = (\vec{p}_{t, i}, v)$, closer to the best job selection decisions.

\begin{figure}[!h]
    \centering
    
    \begin{subfigure}[b]{\columnwidth}
        \includegraphics[width=1\columnwidth]{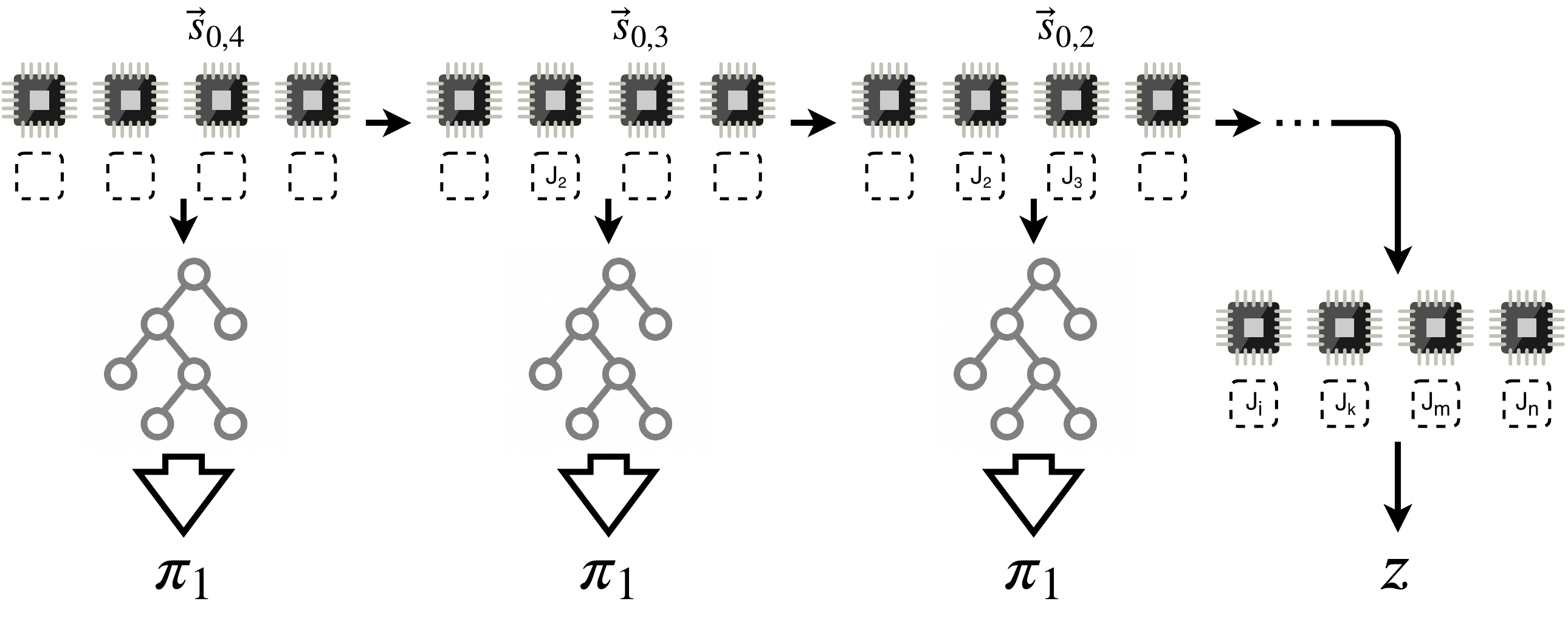}
        \caption{Simulation Procedure.}
    \end{subfigure}
    
    \begin{subfigure}[b]{\columnwidth}
        \includegraphics[width=1\columnwidth]{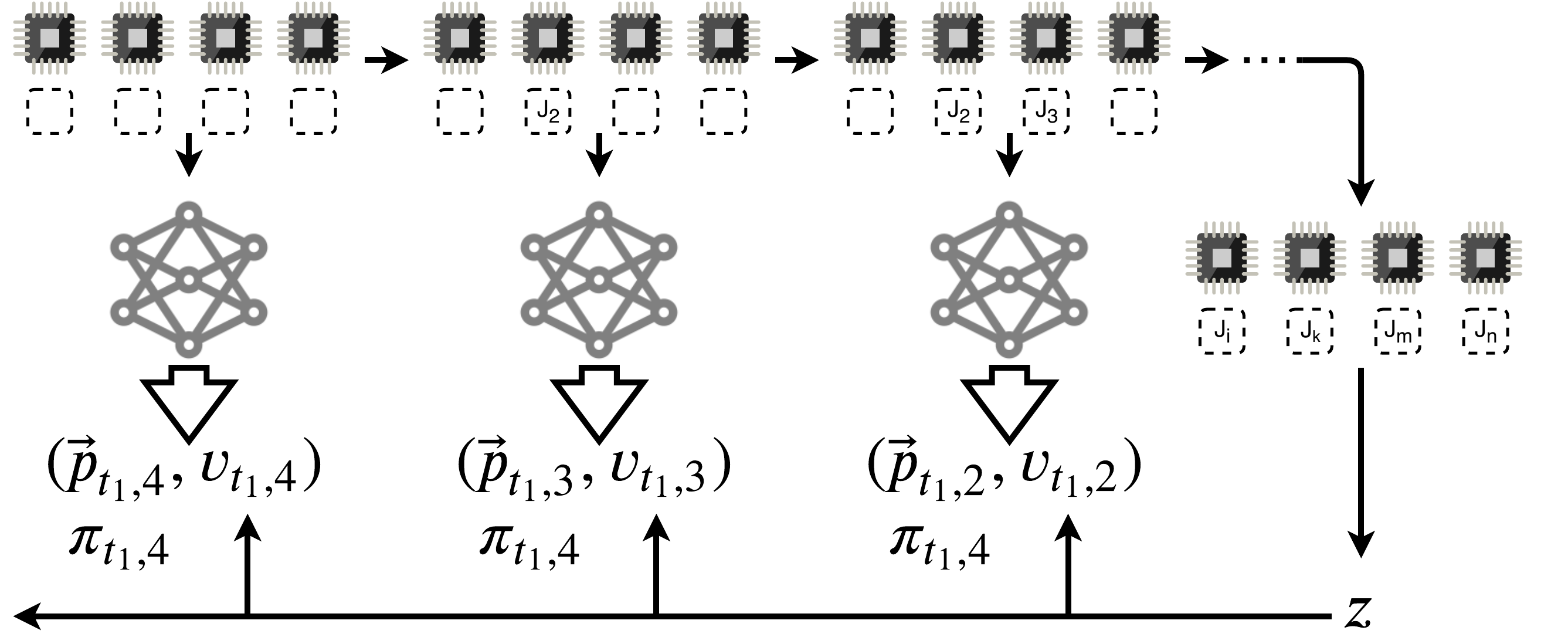}
        \caption{Neural Network Training.}
    \end{subfigure}
    
    % \subfloat[Simulation Procedure.]{\includegraphics[width=1\columnwidth]{figures/simulation.pdf}}

    % \subfloat[Neural Network Training. ]{\includegraphics[width=1\columnwidth]{figures/iteration.pdf}}
    
    \caption{(a) Policy iteration. The simulator do the simulations from $\vec{s}_{0, 4}$ to $\vec{s}_T$. For each system state $\vec{s}_{t, i}$, an MCTS is executed using the deep neural network with the latest parameters $\theta$. Job selections are determined through the probabilities $\pi$ computed by the MCTS. The simulation terminates while any job misses its deadline or a feasible schedule is found (the simulation lasts for a whole hyper-period). The terminal system state $\vec{s}_T$ is evaluated by Equation~\ref{eq:value_update}. (b) Neural network training. The neural network $f_{\theta}$ takes the system state $\vec{s}_{t, i}$ as its input and output both a probabilities tensor $\vec{p}_{t, i}$ and a evaluation value $v_{t, i}$. $\vec{p}_{t, i}$ represents the probability distribution over job selections, and $v_{t, i}$ reflects how well the system works. The parameter of the neural network $\theta$ will be updated to minimize the errors between the search probabilities $\pi_{t, i}$, and to minimize the error between the output evaluation value $v_{t, i}$ with the actual evaluation value. The updated parameters will be used in the next iteration.}
    \label{fig:policy_iteration}
\end{figure}

Fig.~\ref{fig:policy_iteration} illustrates how MCTS computes the optimal State-Action mapping $\vec{\pi}$, starting from $\vec{\pi_0}$:
\begin{itemize}
    \item \textbf{Policy evaluation.} Use simulator to get the value of policy $\vec{\pi}_0$, $v_{\vec{\pi}_0}$.
    \item \textbf{Policy improvement.} Optimize the policy to $\vec{\pi}_1$ according to the evaluation value $v_{\vec{\pi}_0}$. 
    \item Repeat step 1 and step 2 until the optimal policy $\vec{\pi}$ is found.
\end{itemize}

In order to keep the size of the input of the DNN constant, we do zero-padding or truncate the state tensor $\vec{s}_t$ according to the absolute deadlines $D$. 

The DNN is trained by the policy iteration training procedure for each job selection (note that all the job selections done in the MCTS are only simulated). The steps to train the neural network are:
\begin{itemize}
    \item The DNN is initialized to random parameters, $\theta_0$.
    \item At each subsequent iteration, $k \geq 1$, do a scheduling simulation (shown in Fig.~\ref{fig:policy_iteration}). 
    \item For each iteration (at time point $t$) and for each system state $\vec{s}_{t, i}$, a Monte-Carlo Tree Search is executed using the previous iteration of DNN $f_{\theta_{k-1}}$. The job selection is done by maximizing the upper confidence bound for $Q(\vec{s}_{t, i}, \vec{a}_{t, i}) + U(\vec{s}_{t, i}, \vec{a}_{t, i})$.
    \item A simulation stops/terminates at time point $t'$, when any job misses its deadline or $t' > H(\tau)$. The simulation is than scored (Equation~\ref{eq:value_update}) to calculate the reward $z_i$. 
    \item The data for each system state is stored as $(\vec{s}_{t, i}, \vec{\pi}_{t, i}, z_{t, i})$, and the parameters for DNN ($\theta_{k-1}$) are updated to $\theta_{k}$  using the data uniformly sampled from all system states of the last iteration by using $(\vec{s}, \vec{\pi}, z)$ as the training samples. 
\end{itemize}

The training objective for the DNN is to minimize the error between $v_{t, i}$ and  $z_{t, i}$, and to maximize the similarity between $\vec{p}_{t, i}$ and $\vec{\pi}_{t, i}$. Thus, the loss function used by the gradient descent procedure to adjust the parameters $\theta$ of the DNN can be written as:
\begin{equation}
    (\vec{p}, v) = f_\theta(\vec{s}), l = |z - v| - \vec{\pi}^T \log \vec{p} + c\|\theta\|^2
    \label{eq:loss}
\end{equation}

% \subsubsection{Minimizing Overhead}
% We also minimize overheads due to DNN decision making. Since we plan to use DNN as a dynamic runtime scheduler, the latency overhead due to DNN evaluation and decision making shall be kept minimum, constraining us from having an enormous DNN structure. Hence, there is a tradeoff between the expressibility vs. overhead of the DNN, which we have to take into consideration while designing the DNN's structure. Hence the chosen DNN structure is small compared to the DNNs used for other applications such as Go, object detectors, yet still being sufficient in yielding competitive performance as proven by evaluation. 

\subsubsection{Decision Phase for Dynamic Configuration}
\label{sec:decision_dynamic}

As discussed earlier, we use the DNN as a dynamic runtime scheduler. Given the current state $\vec{s}_{t,m}$ of the system we get the most probable action by finding $i = max\_index(\vec{p})$ where $f_{\theta}(\vec{s}_{t,m}) = (\vec{p},v)$. We mark $J_i$ to be executed in processor 1 for the next time slot and update the state tensor $\vec{s}_{t,m-1}$ by changing only the entries corresponding to $J_i$. Then, $\vec{s}_{t,m-1}$ is given as the input to the DNN and the same procedure repeats until we reach $\vec{s}_{t,1}$ or some $\vec{s}_{t,i}$ which is empty. Jobs that have been marked for execution are sent to their corresponding processors by performing context switches, or migrations wherever required, and the system resumes execution for one time slot. After this timeslot, we get the new state tensor $\vec{s}_{t+1,m}$, and the above procedure is repeated.

\subsubsection{Decision Phase for Static Configuration}
\label{sec:decision_static}

For this configuration, we are given a static task set,  RL training starts from $\vec{s}_{0,m}$. We use the DNN guided MCTS to performs roll-outs. After one roll-out is complete, if we find a feasible schedule, then we return that schedule. Otherwise, we compute the reward and train the DNN according to that reward. This procedure is repeated for subsequent roll-outs until either a roll-out manages to find a feasible schedule or the roll-out threshold is reached. If the roll-out threshold is reached, our algorithm has failed to find a feasible schedule. In that case, we can re-run the algorithm with a higher roll-out threshold.

\subsection{Extensions}
\label{sec:extensions}
In addition to preemptive multiprocessor systems, \name{} can be easily extended to handle other complexities 
% seen in both task and hardware models. We demonstrate herein how to extend \name{} to support 
such as non-preemptive systems and/or heterogeneous multiprocessors. The flexibility and extensibility of \name{} are fundamentally due to the fact that it is a training-based approach. Thus, one only needs to modify the system state tensor when facing complexities in the models. The DNN can automatically learn to convert this new state to a reasonable action and, in turn, guide the MCTS to regions that conform to these new constraints.

\subsubsection{Non-preemptive System}
For non-preemptive workloads, the system state tensor has the same format as the tensor in Sec.~\ref{eq:preemptive_system_state}. Since we  do not need to consider the running jobs for selection, the state tensor here is derived only from the job waiting list of the system, and the parameter $i$ starts from the number of idle processors. 

% The following is an example of the system state:
% \begin{equation}
%     \label{eq:non-preemptive_system_state}
%     \vec{s}_{t, i} = 
%     \begin{bmatrix}
%         D_{i_1}^{(j_1)} - t & E_{i_1}^{(j_1)} - \mathrm{execution}(R_{i_1}^{(j_1)}, t) & 0  \\
%         D_{i_2}^{(j_2)} - t & E_{i_2}^{(j_2)} - \mathrm{execution}(R_{i_2}^{(j_2)}, t) & 0  \\
%         \vdots & \vdots \\
%         D_{i_n}^{(j_n)} - t & E_{i_n}^{(j_n)} - \mathrm{execution}(R_{i_n}^{(j_n)}, t) & 0
%     \end{bmatrix}
% \end{equation}

\subsubsection{Heterogeneous Multiprocessors}
For heterogeneous multiprocessors, $\vec{s}_{t, i}$ has the same format. However, since the job will have different execution times on different types of processors, the second column of $\vec{s}_{t, i}$ needs modification.
% we need to modify the remaining execution time column of the system state according to the processor type.
For example, assuming there are two types of processors in the system, a computation that needs to consume 1 millisecond on a processor of type 1 will need 2 milliseconds to compute on a processor of type 2 because a processor of type 1 is twice as fast. First, we define the execution of $J_i^{(j)}$ at time $t$ as follows:
$$\textrm{execution}(R_i^{(j)}, t) = \textrm{execution}(R_i^{(j)}, t, 1) + \frac{\textrm{execution}(R_i^{(j)}, t, 2)}{2}  $$
where $\textrm{execution}(R_i^{(j)}, t, k)$ denotes the time allocated on processors of type $k$ for $J_i^{(j)}$ in the window $[R_i^{(j)},t)$. Thus, the system state tensor that is used for selecting a job to run on a type 1 processor is same as the tensor for the homogeneous case, 
% \begin{equation}
%     \label{eq:heterogeneous_system_state_1}
%     \vec{s}_{t, i} ^{(1)} = 
%     \begin{bmatrix}
%         D_{i_1}^{(j_1)} - t & E_{i_1}^{(j_1)} - \mathrm{execution}(R_{i_1}^{(j_1)}, t) & 0/1  \\
%         D_{i_2}^{(j_2)} - t & E_{i_2}^{(j_2)} - \mathrm{execution}(R_{i_2}^{(j_2)}, t) & 0/1  \\
%         \vdots & \vdots \\
%         D_{i_n}^{(j_n)} - t & E_{i_n}^{(j_n)} - \mathrm{execution}(R_{i_n}^{(j_n)}, t) & 0/1
%     \end{bmatrix}, 
% \end{equation}
whereas  the system state used for a type 2 processor is:
\begin{equation}
    \label{eq:heterogeneous_system_state_2}
    \vec{s}_{t, i} ^{(2)} = 
    \begin{bmatrix}
        D_{i_1}^{(j_1)} - t & 2 \times [E_{i_1}^{(j_1)} - \mathrm{execution}(R_{i_1}^{(j_1)}, t)] & 0/1  \\
        D_{i_2}^{(j_2)} - t & 2 \times [E_{i_2}^{(j_2)} - \mathrm{execution}(R_{i_2}^{(j_2)}, t)] & 0/1  \\
        \vdots & \vdots \\
        D_{i_n}^{(j_n)} - t & 2 \times [E_{i_n}^{(j_n)} - \mathrm{execution}(R_{i_n}^{(j_n)}, t)] & 0/1
    \end{bmatrix}
\end{equation}
We first select jobs to execute on type 1 processors and then construct the next intermediate state for type 2 processors for selecting jobs to execute on them. Note that the scheduling overhead is similar to the homogeneous case as the number of intermediary states is the same, and the conversion between the states for type 1 and 2 processors is trivial.

\subsubsection{Scheduling-related Overhead Consideration} 
Note that \name{} takes into account the various kinds of overheads due to the scheduling process in the design space.
Similar to the above-mentioned extensions, the base workflow remains the same. The only needed modification is on the system state representation. We rely on an existing straightforward overhead accounting method~\cite{brandenburg2009accounting} and gather the actual overheads for running task sets in LITMUS$^{RT}$, which is a Linux-based testbed for evaluating various scheduling algorithms. The system state considering such overheads modifies each job's worst-case execution time from $E$ to
% \begin{equation}
%     \label{eq:state_with_overheads}
%     \vec{s}_{t, i} = 
%     \begin{bmatrix}
%         D_{i_1}^{(j_1)} - t & {E'}_{i_1}^{(j_1)} - \mathrm{execution}(R_{i_1}^{(j_1)}, t) & 0/1  \\
%         D_{i_2}^{(j_2)} - t & {E'}_{i_2}^{(j_2)} - \mathrm{execution}(R_{i_2}^{(j_2)}, t) & 0/1  \\
%         \vdots & \vdots \\
%         D_{i_n}^{(j_n)} - t & {E'}_{i_n}^{(j_n)} - \mathrm{execution}(R_{i_n}^{(j_n)}, t) & 0/1
%     \end{bmatrix}
% \end{equation}
$E' = \bar{o} + E.$ 
Where $\bar{o}$ denotes the average overhead for all the tasks in $\tau$, which is an empirical value originating from the data collected in LITMUS$^{RT}$.

\begin{figure*}[h]
    \captionsetup[subfigure]{aboveskip=-4pt,belowskip=-3pt}
    \centering
    
    % \begin{subfigure}[b]{0.24\textwidth}
    %     \includegraphics[width=\textwidth]{figures/results/preemptive_homo_2_light.pdf}
    %     \caption{$m=2$, light.}
    % \end{subfigure}
    % \begin{subfigure}[b]{0.24\textwidth}
    %     \includegraphics[width=\textwidth]{figures/results/preemptive_homo_2_medium.pdf}
    %     \caption{$m=2$, medium.}
    % \end{subfigure}
    % \begin{subfigure}[b]{0.24\textwidth}
    %     \includegraphics[width=\textwidth]{figures/results/preemptive_homo_2_heavy.pdf}
    %     \caption{$m=2$, heavy.}
    % \end{subfigure}
    % \begin{subfigure}[b]{0.24\textwidth}
    %     \includegraphics[width=\textwidth]{figures/results/preemptive_homo_2_mixed.pdf}
    %     \caption{$m=2$, mixed.}
    % \end{subfigure}
    
    \begin{subfigure}[b]{0.24\textwidth}
        \includegraphics[width=\textwidth]{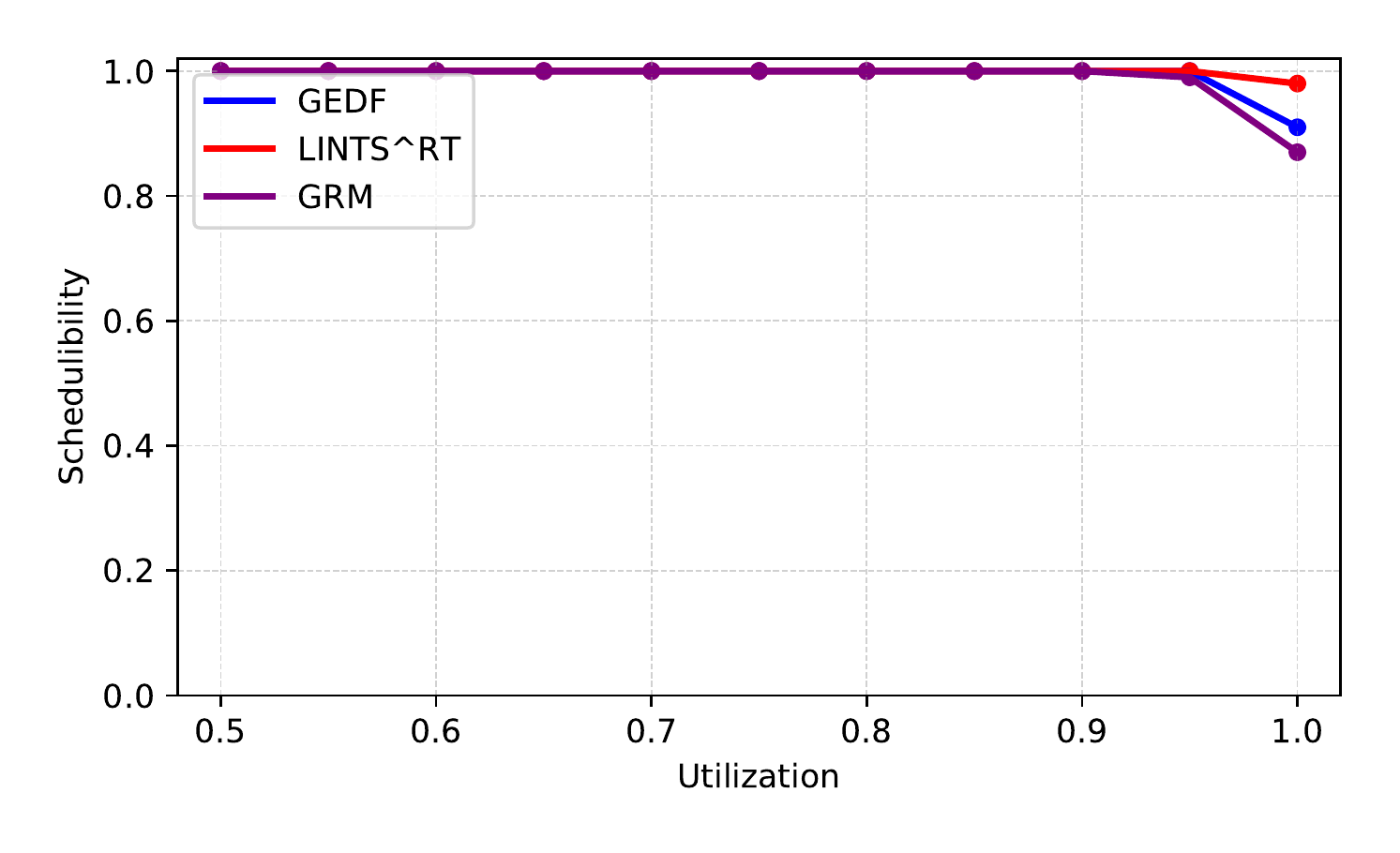}
        \vspace{-4mm}
        \caption{light.}
    \end{subfigure}
    \begin{subfigure}[b]{0.24\textwidth}
        \includegraphics[width=\textwidth]{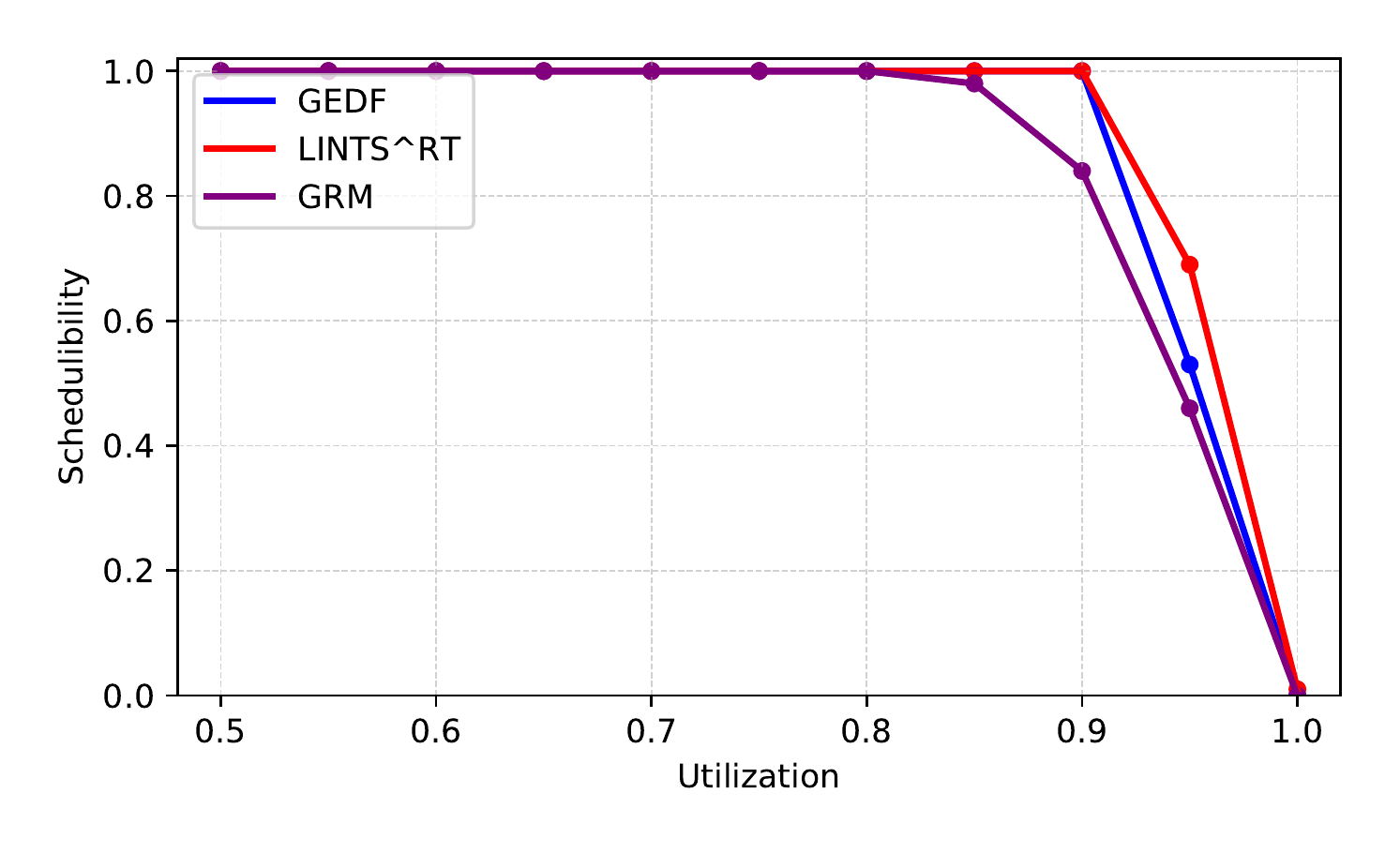}
           \vspace{-4mm}
        \caption{medium.}
    \end{subfigure}
    \begin{subfigure}[b]{0.24\textwidth}
        \includegraphics[width=\textwidth]{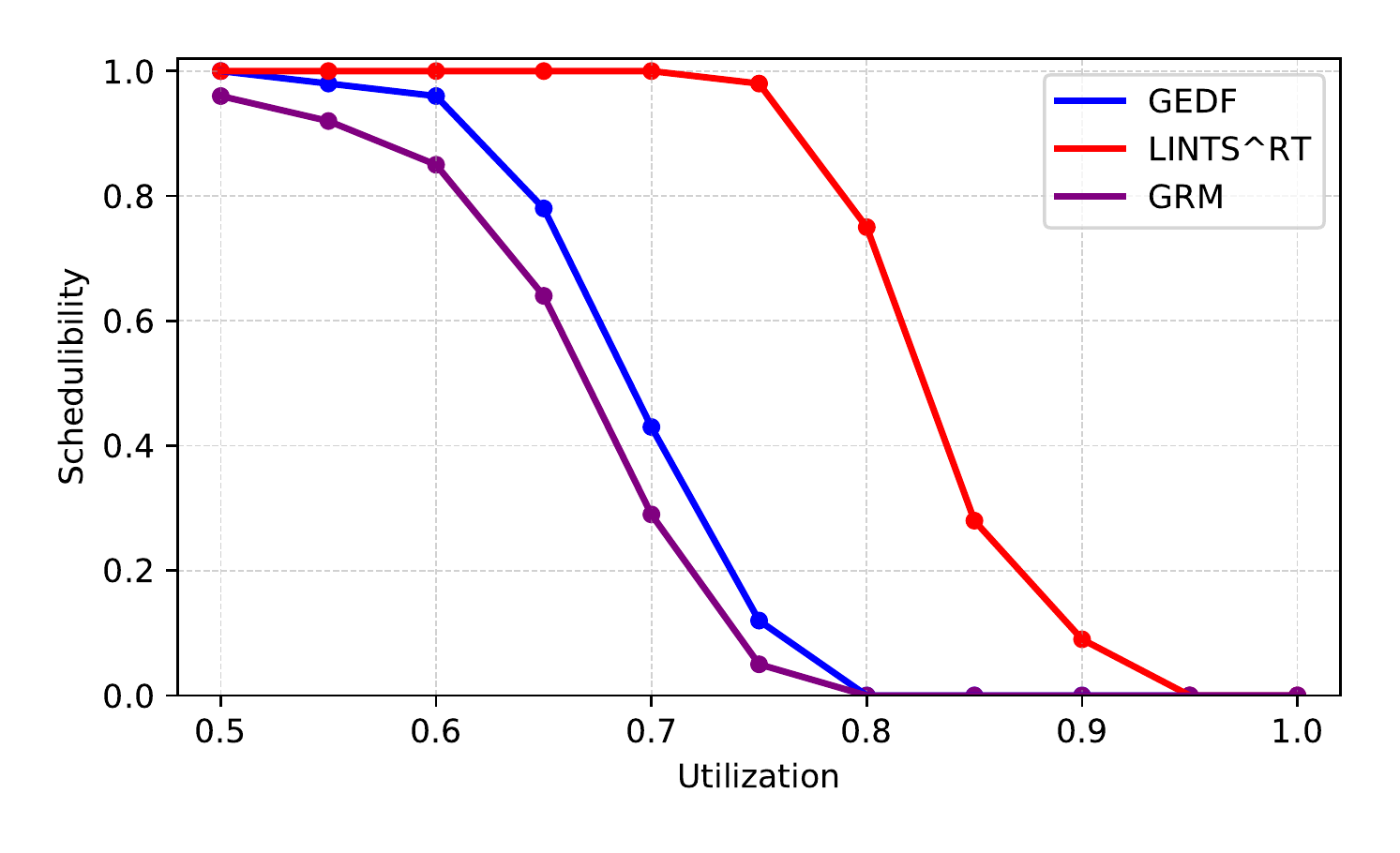}
           \vspace{-4mm}
        \caption{heavy.}
    \end{subfigure}
    \begin{subfigure}[b]{0.24\textwidth}
        \includegraphics[width=\textwidth]{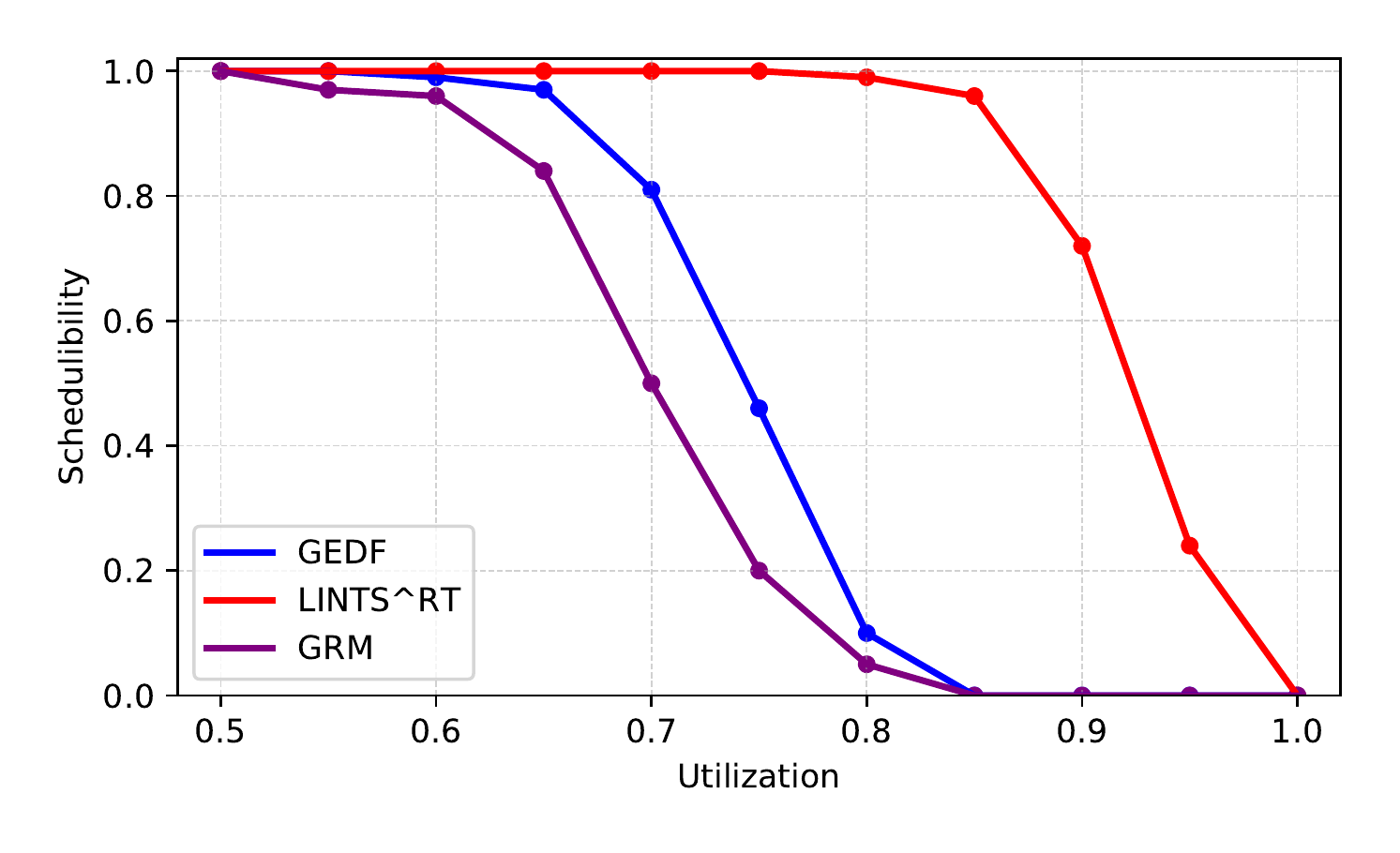}
           \vspace{-4mm}
        \caption{mixed.}
    \end{subfigure}
    
    % \subfloat[Per-task-utilization: light.]{\includegraphics[width=0.24\textwidth]{figures/results/preemptive_homo_4_light.pdf}}
    % \subfloat[Per-task-utilization: medium.]{\includegraphics[width=0.24\textwidth]{figures/results/preemptive_homo_4_medium.pdf}}
    % \subfloat[Per-task-utilization: heavy.]{\includegraphics[width=0.24\textwidth]{figures/results/preemptive_homo_4_heavy.pdf}}
    % \subfloat[Per-task-utilization: mixed.]{\includegraphics[width=0.24\textwidth]{figures/results/preemptive_homo_4_mixed.pdf}}

    % \subfloat[Per-task-utilization: light.]{\includegraphics[width=0.24\textwidth]{figures/results/preemptive_homo_2_light.pdf}}
    % \subfloat[Per-task-utilization: medium.]{\includegraphics[width=0.24\textwidth]{figures/results/preemptive_homo_2_medium.pdf}}
    % \subfloat[Per-task-utilization: heavy.]{\includegraphics[width=0.24\textwidth]{figures/results/preemptive_homo_2_heavy.pdf}}
    % \subfloat[Per-task-utilization: mixed.]{\includegraphics[width=0.24\textwidth]{figures/results/preemptive_homo_2_mixed.pdf}}
    
    \caption{Runtime schedulbility for ($m=4$, Preemptive, homogeneous, \{light, medium, heavy, mixed\}) vs. $\frac{U}{m}$ }
    \label{fig:preemptive_homogeneous}
\end{figure*}

\begin{figure*}[t]
    \captionsetup[subfigure]{aboveskip=-4pt,belowskip=-3pt}
    \centering
    %     \begin{subfigure}[b]{0.24\textwidth}
    %     \includegraphics[width=\textwidth]{figures/results/preemptive_hetero_2_light.pdf}
    %     \caption{$m=2$, light.}
    % \end{subfigure}
    % \begin{subfigure}[b]{0.24\textwidth}
    %     \includegraphics[width=\textwidth]{figures/results/preemptive_hetero_2_medium.pdf}
    %     \caption{$m=2$, medium.}
    % \end{subfigure}
    % \begin{subfigure}[b]{0.24\textwidth}
    %     \includegraphics[width=\textwidth]{figures/results/preemptive_hetero_2_heavy.pdf}
    %     \caption{$m=2$, heavy.}
    % \end{subfigure}
    % \begin{subfigure}[b]{0.24\textwidth}
    %     \includegraphics[width=\textwidth]{figures/results/preemptive_hetero_2_mixed.pdf}
    %     \caption{$m=2$, mixed.}
    % \end{subfigure}

        \begin{subfigure}[b]{0.24\textwidth}
        \includegraphics[width=\textwidth]{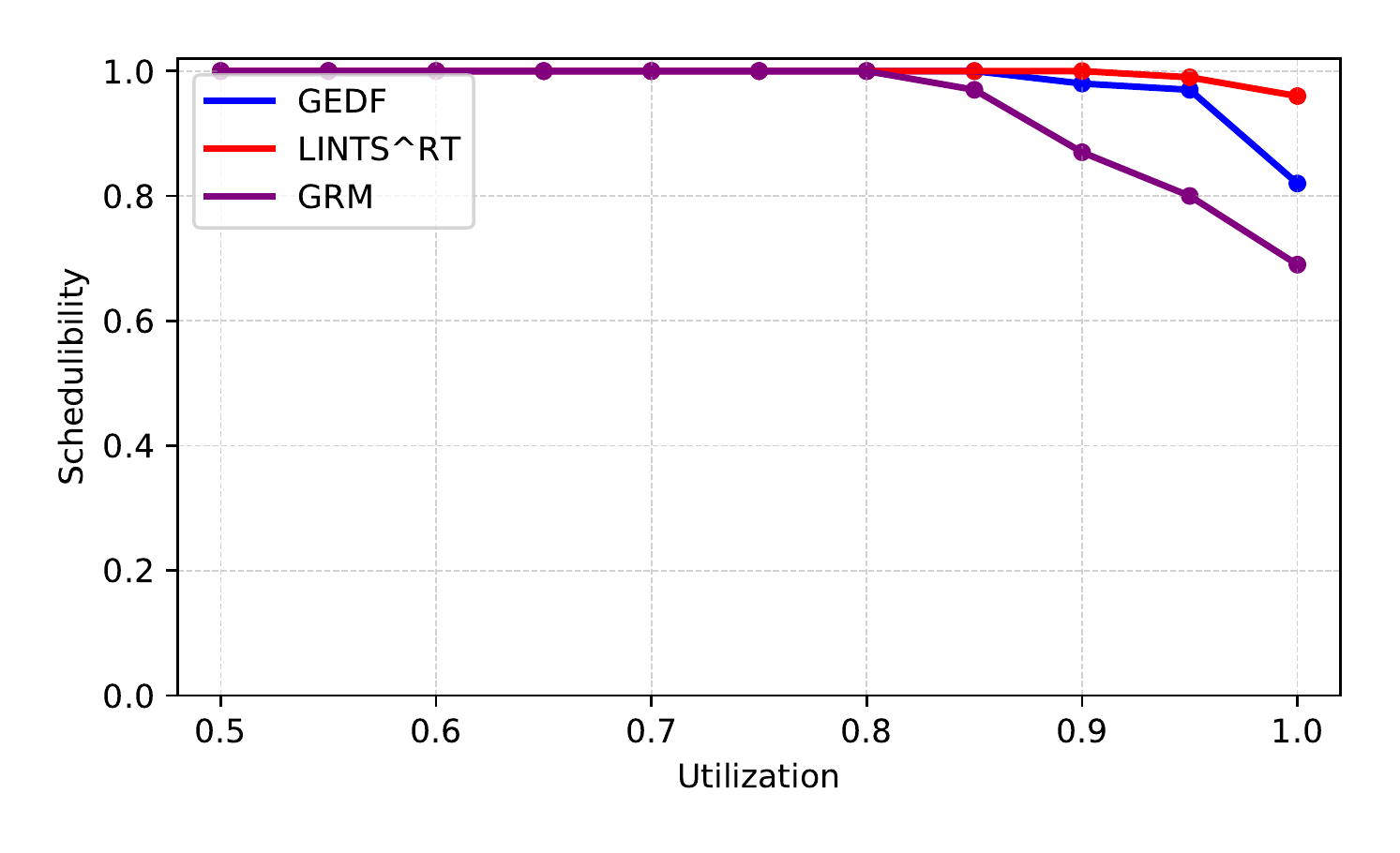}
        \caption{light.}
    \end{subfigure}
    \begin{subfigure}[b]{0.24\textwidth}
        \includegraphics[width=\textwidth]{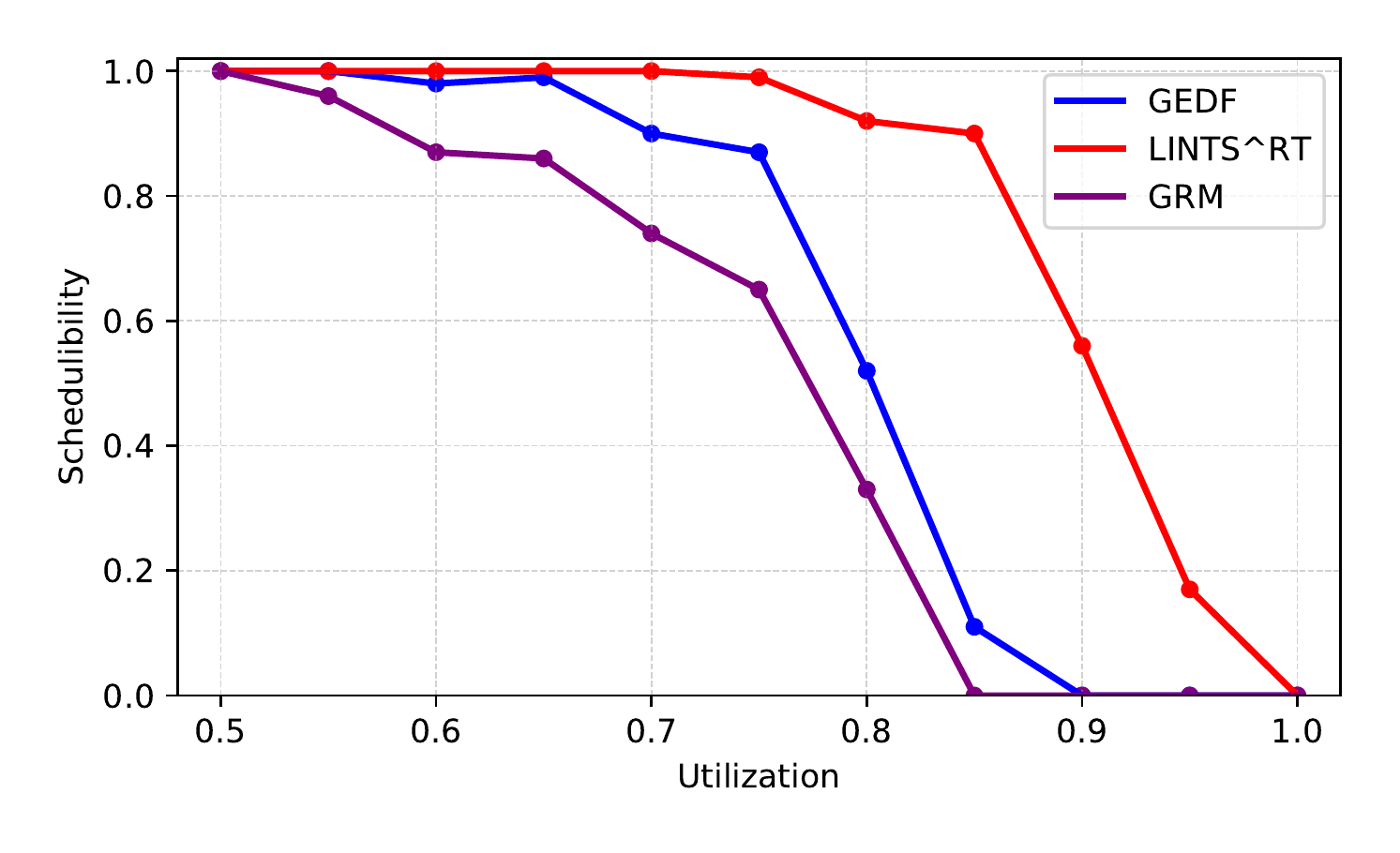}
        \caption{medium.}
    \end{subfigure}
    \begin{subfigure}[b]{0.24\textwidth}
        \includegraphics[width=\textwidth]{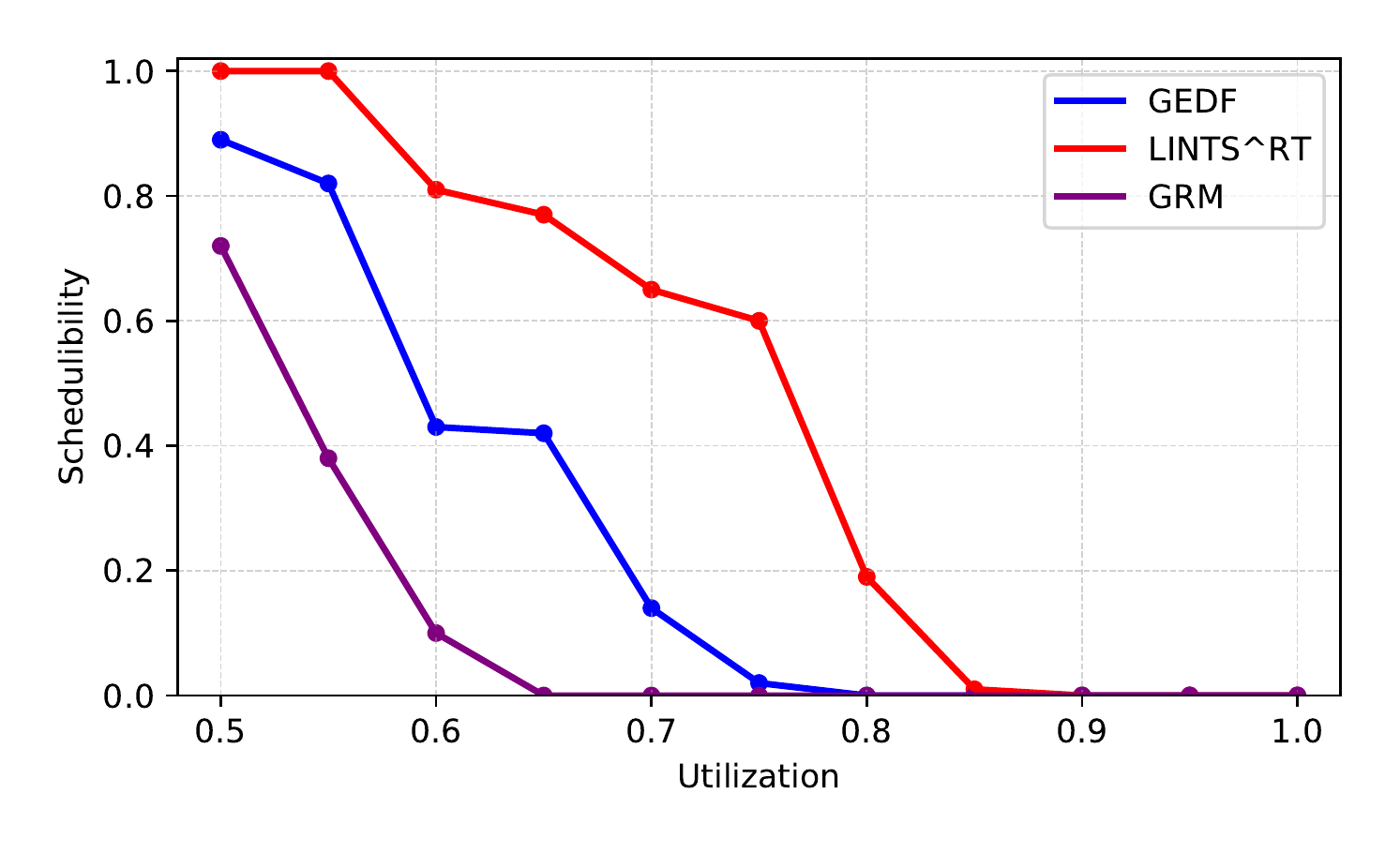}
        \caption{heavy.}
    \end{subfigure}
    \begin{subfigure}[b]{0.24\textwidth}
        \includegraphics[width=\textwidth]{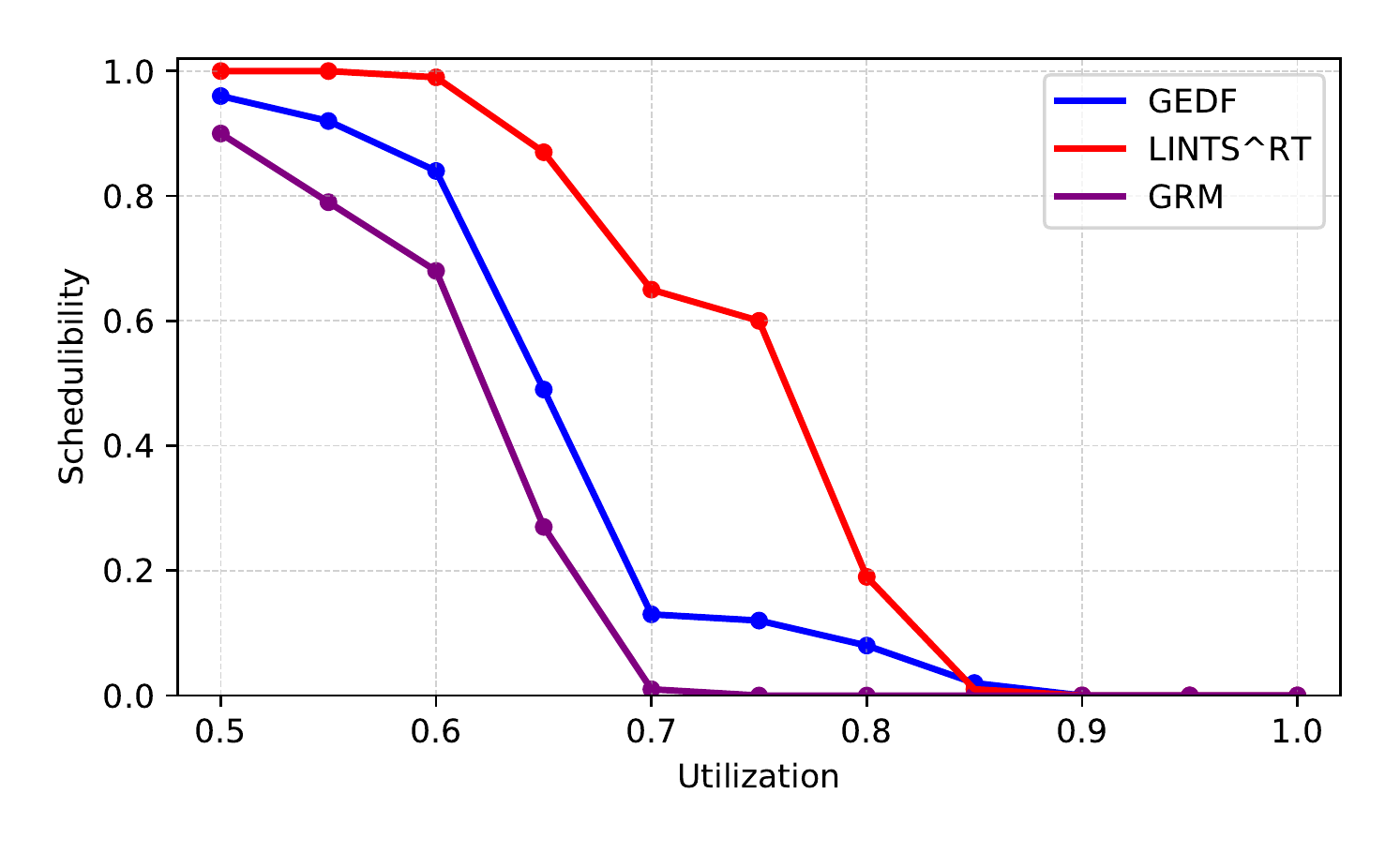}
        \caption{mixed.}
    \end{subfigure}
    
    % \subfloat[Per-task-utilization: light.]{\includegraphics[width=0.24\textwidth]{figures/results/preemptive_hetero_4_light.pdf}}
    % \subfloat[Per-task-utilization: medium.]{\includegraphics[width=0.24\textwidth]{figures/results/preemptive_hetero_4_medium.pdf}}
    % \subfloat[Per-task-utilization: heavy.]{\includegraphics[width=0.24\textwidth]{figures/results/preemptive_hetero_4_heavy.pdf}}
    % \subfloat[Per-task-utilization: mixed.]{\includegraphics[width=0.24\textwidth]{figures/results/preemptive_hetero_4_mixed.pdf}}

    % \subfloat[Per-task-utilization: light.]{\includegraphics[width=0.24\textwidth]{figures/results/preemptive_hetero_2_light.pdf}}
    % \subfloat[Per-task-utilization: medium.]{\includegraphics[width=0.24\textwidth]{figures/results/preemptive_hetero_2_medium.pdf}}
    % \subfloat[Per-task-utilization: heavy.]{\includegraphics[width=0.24\textwidth]{figures/results/preemptive_hetero_2_heavy.pdf}}
    % \subfloat[Per-task-utilization: mixed.]{\includegraphics[width=0.24\textwidth]{figures/results/preemptive_hetero_2_mixed.pdf}}
    
    \caption{Runtime schedulbility for ($m=4$,  Preemptive, heterogeneous, \{light, medium, heavy, mixed\}) vs. $\frac{U}{m}$ }
    \label{fig:preemptive_heterogeneous_8}
\end{figure*}
\begin{figure*}[t]
    \captionsetup[subfigure]{aboveskip=-4pt,belowskip=-3pt}
    \centering
    
    % \begin{subfigure}[b]{0.24\textwidth}
    %     \includegraphics[width=\textwidth]{figures/results/non-preemptive_homo_2_light.pdf}
    %     \caption{$m=2$, light.}
    % \end{subfigure}
    % \begin{subfigure}[b]{0.24\textwidth}
    %     \includegraphics[width=\textwidth]{figures/results/non-preemptive_homo_2_medium.pdf}
    %     \caption{$m=2$, medium.}
    % \end{subfigure}
    % \begin{subfigure}[b]{0.24\textwidth}
    %     \includegraphics[width=\textwidth]{figures/results/non-preemptive_homo_2_heavy.pdf}
    %     \caption{$m=2$, heavy.}
    % \end{subfigure}
    % \begin{subfigure}[b]{0.24\textwidth}
    %     \includegraphics[width=\textwidth]{figures/results/non-preemptive_homo_2_mixed.pdf}
    %     \caption{$m=2$, mixed.}
    % \end{subfigure}

    \begin{subfigure}[b]{0.24\textwidth}
        \includegraphics[width=\textwidth]{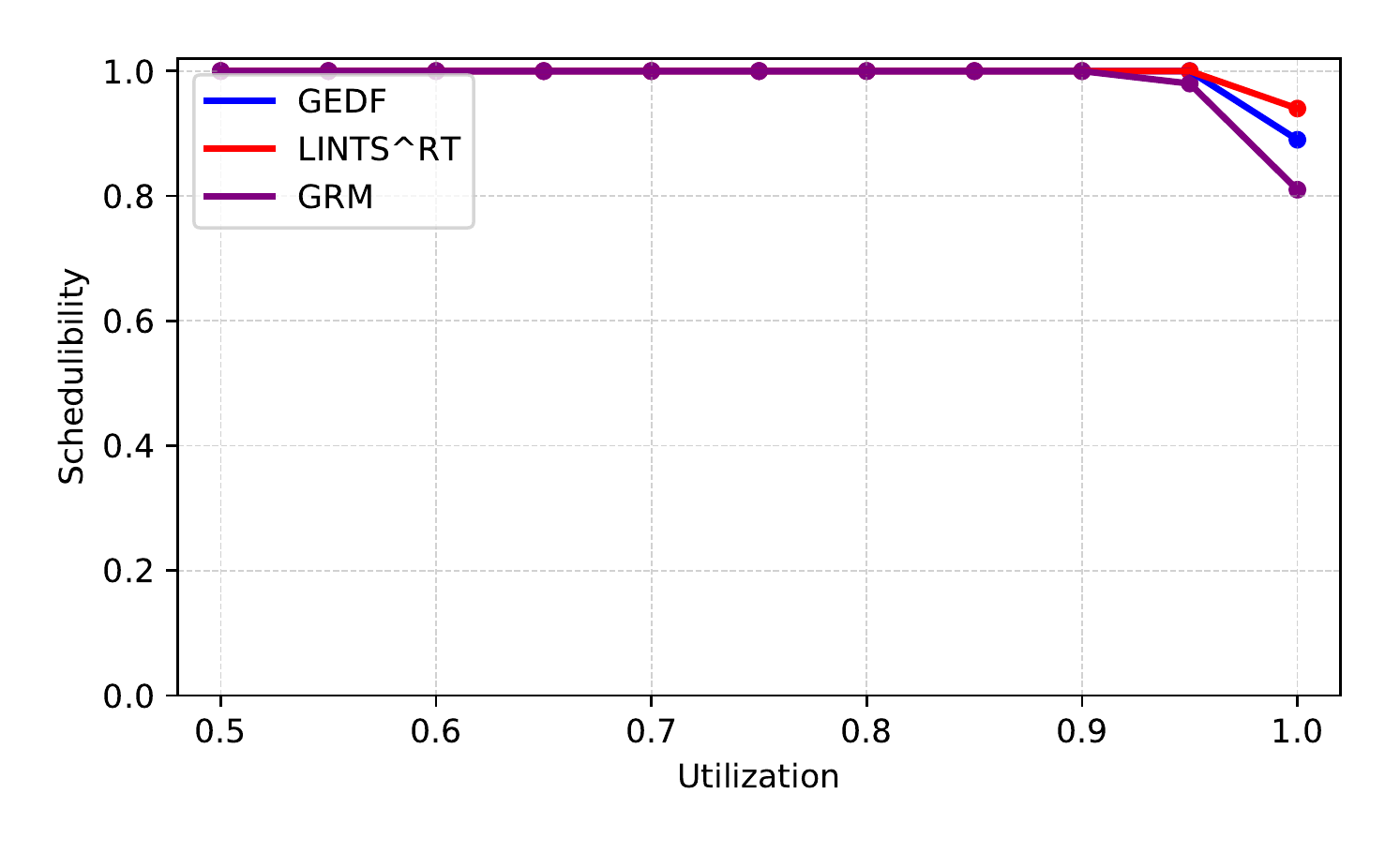}
        % \vspace*{-10mm}
        \caption{light.}
    \end{subfigure}
    \begin{subfigure}[b]{0.24\textwidth}
        \includegraphics[width=\textwidth]{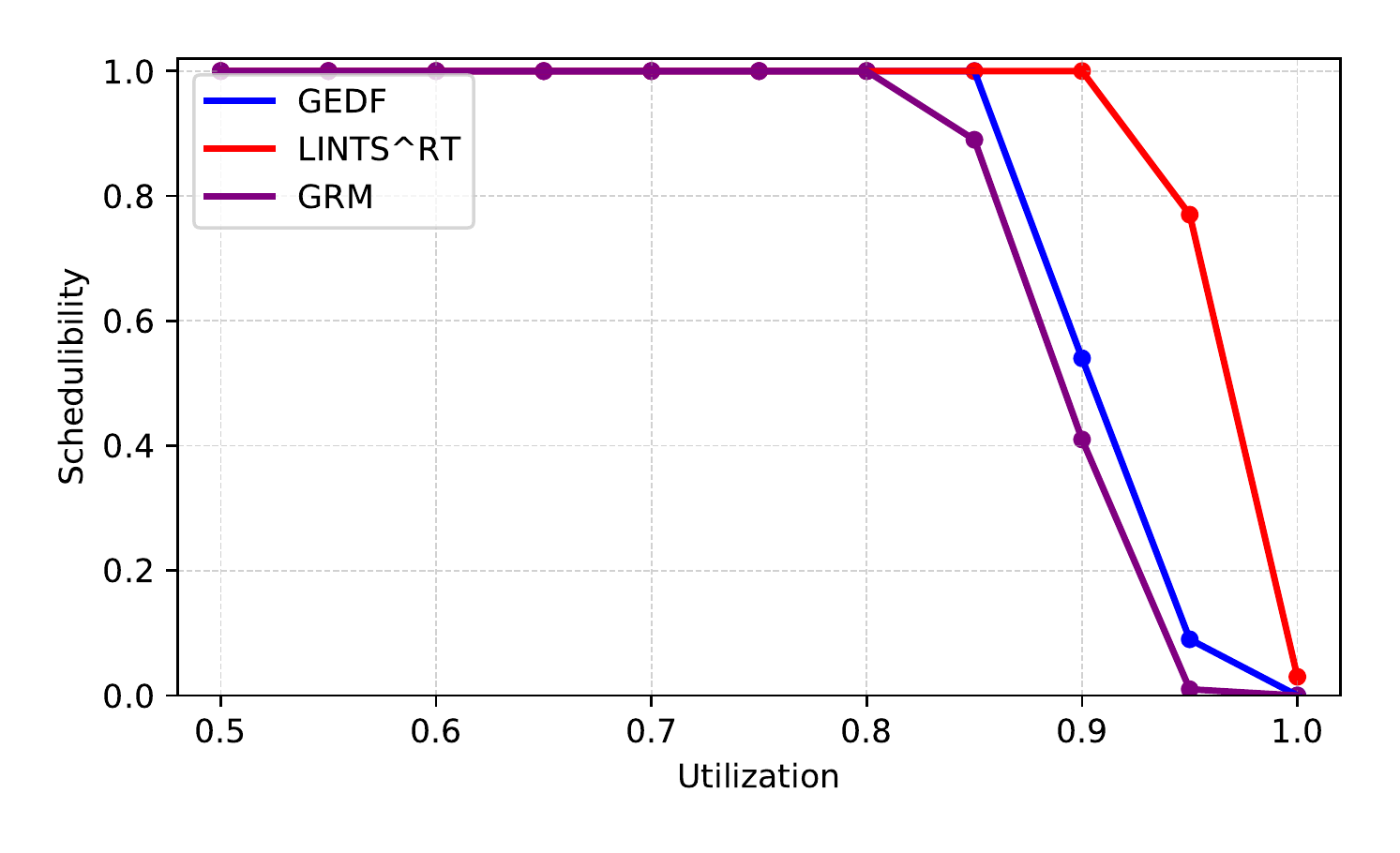}
        \caption{medium.}
    \end{subfigure}
    \begin{subfigure}[b]{0.24\textwidth}
        \includegraphics[width=\textwidth]{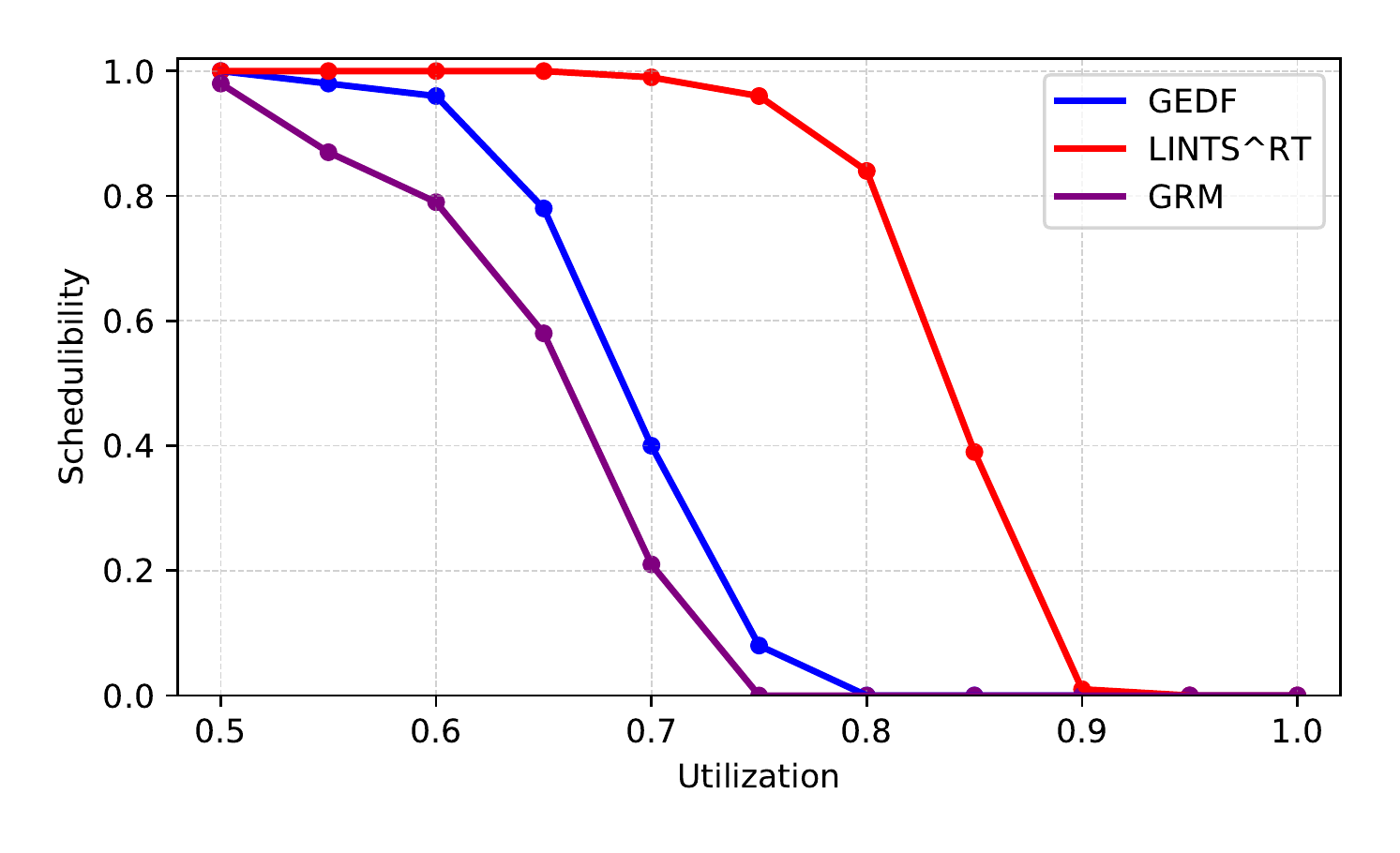}
        \caption{heavy.}
    \end{subfigure}
    \begin{subfigure}[b]{0.24\textwidth}
        \includegraphics[width=\textwidth]{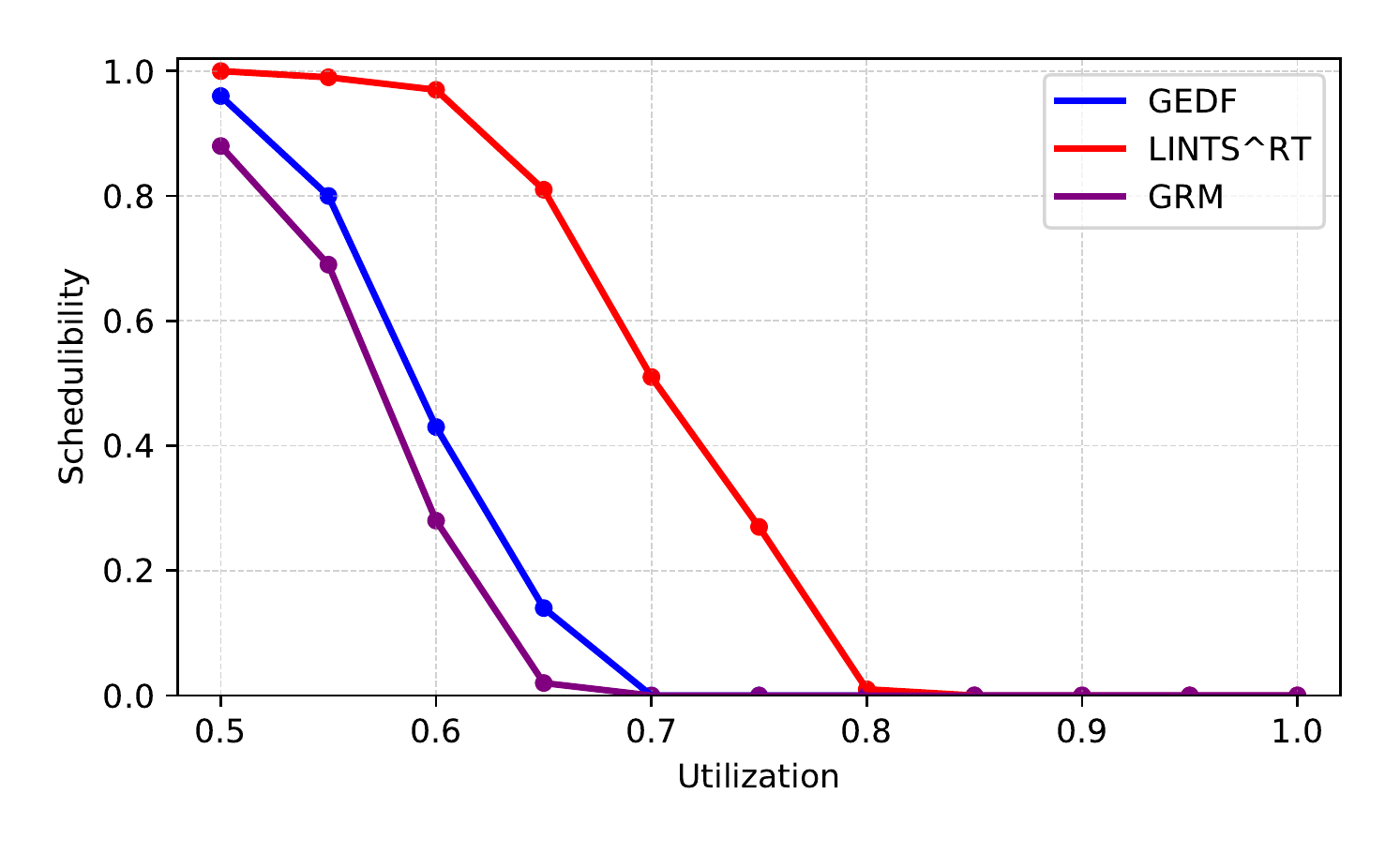}
        \caption{mixed.}
    \end{subfigure}

    \caption{Runtime schedulbility for ($m=4$,  Non-preemptive, homogeneous, \{light, medium, heavy, mixed\}) vs. $\frac{U}{m}$}
    \label{fig:non-preemptive_homogeneous}
\end{figure*}

\begin{figure*}[t]
    \captionsetup[subfigure]{aboveskip=-4pt,belowskip=-3pt}
    \centering
    
    % \begin{subfigure}[b]{0.24\textwidth}
    %     \includegraphics[width=\textwidth]{figures/results/non-preemptive_hetero_2_light.pdf}
    %     \caption{$m=2$, light.}
    % \end{subfigure}
    % \begin{subfigure}[b]{0.24\textwidth}
    %     \includegraphics[width=\textwidth]{figures/results/non-preemptive_hetero_2_medium.pdf}
    %     \caption{$m=2$, medium.}
    % \end{subfigure}
    % \begin{subfigure}[b]{0.24\textwidth}
    %     \includegraphics[width=\textwidth]{figures/results/non-preemptive_hetero_2_heavy.pdf}
    %     \caption{$m=2$, heavy.}
    % \end{subfigure}
    % \begin{subfigure}[b]{0.24\textwidth}
    %     \includegraphics[width=\textwidth]{figures/results/non-preemptive_hetero_2_mixed.pdf}
    %     \caption{$m=2$,  mixed.}
    % \end{subfigure}

    \begin{subfigure}[b!]{0.24\textwidth}
        \includegraphics[width=\textwidth]{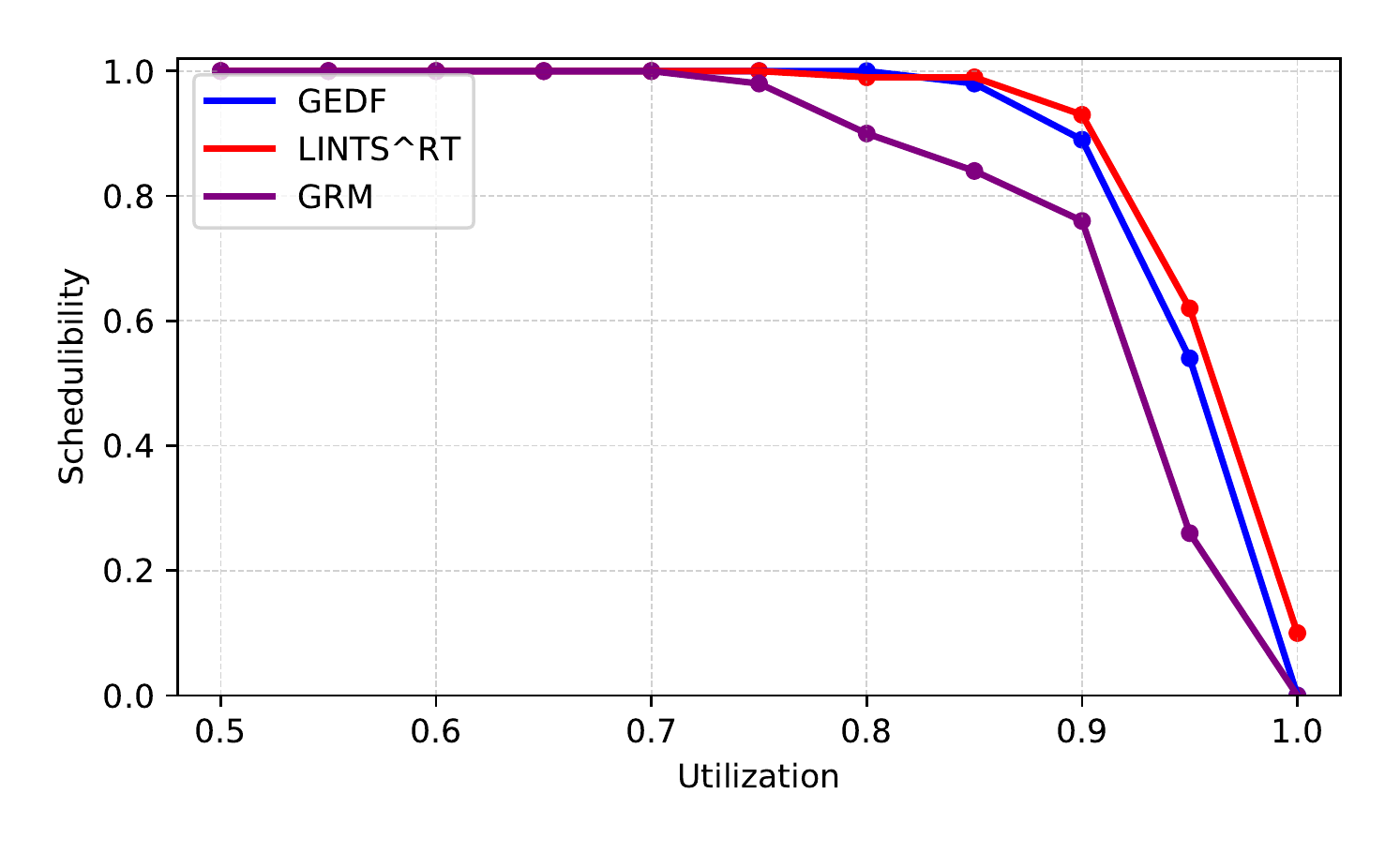}
        \caption{light.}
    \end{subfigure}
    \begin{subfigure}[b!]{0.24\textwidth}
        \includegraphics[width=\textwidth]{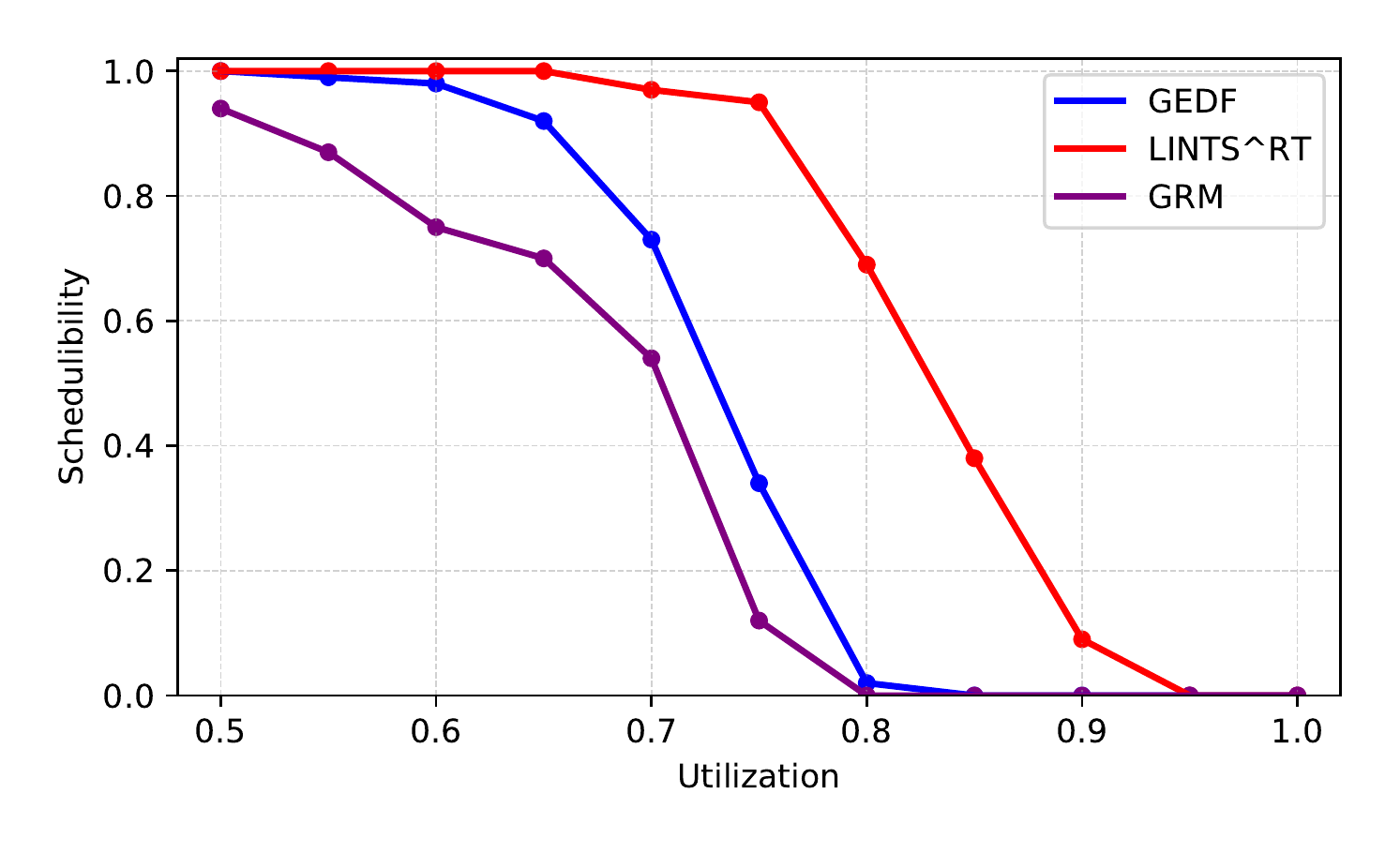}
        \caption{medium.}
    \end{subfigure}
    \begin{subfigure}[b!]{0.24\textwidth}
        \includegraphics[width=\textwidth]{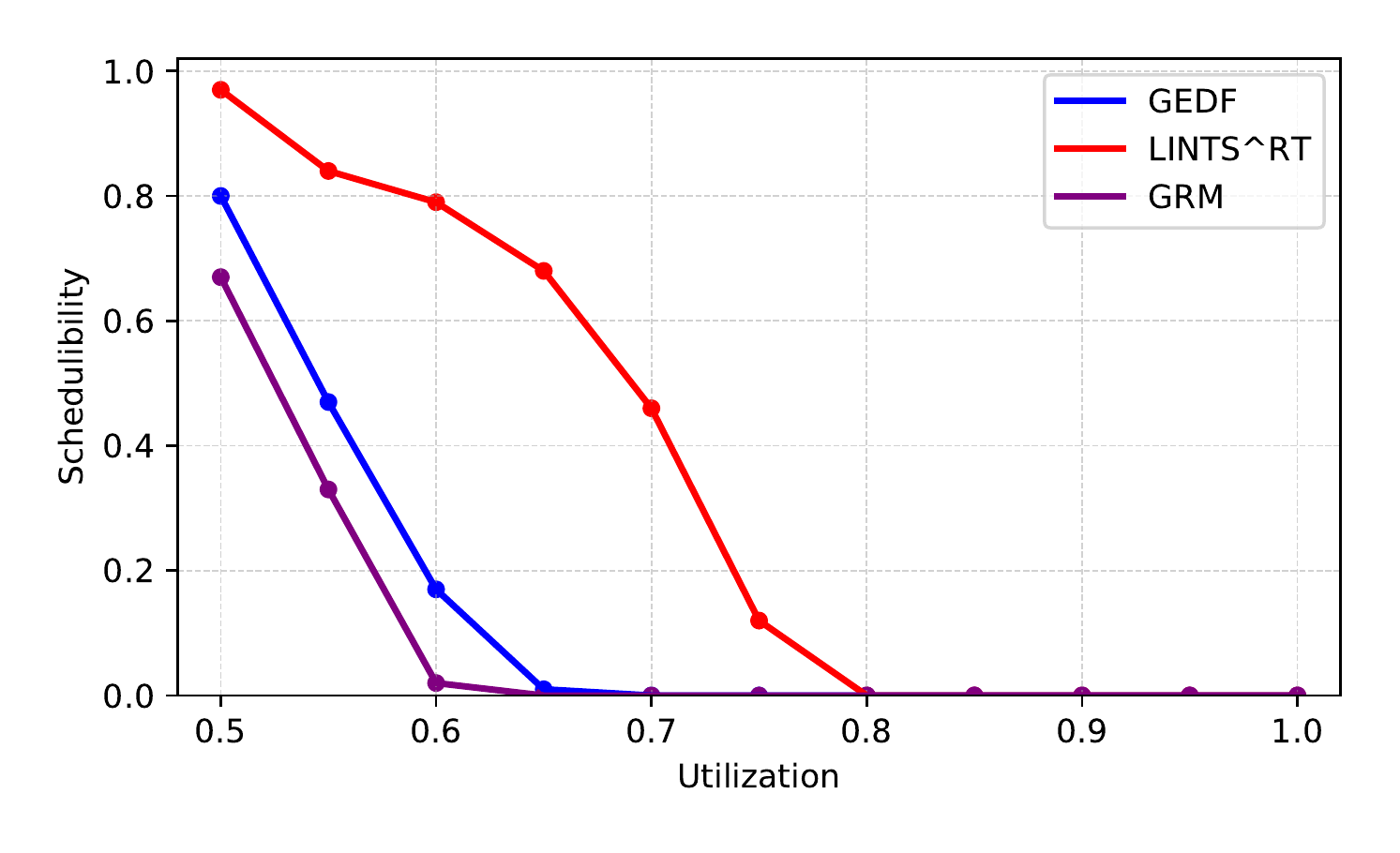}
        \caption{heavy.}
    \end{subfigure}
    \begin{subfigure}[b!]{0.24\textwidth}
        \includegraphics[width=\textwidth]{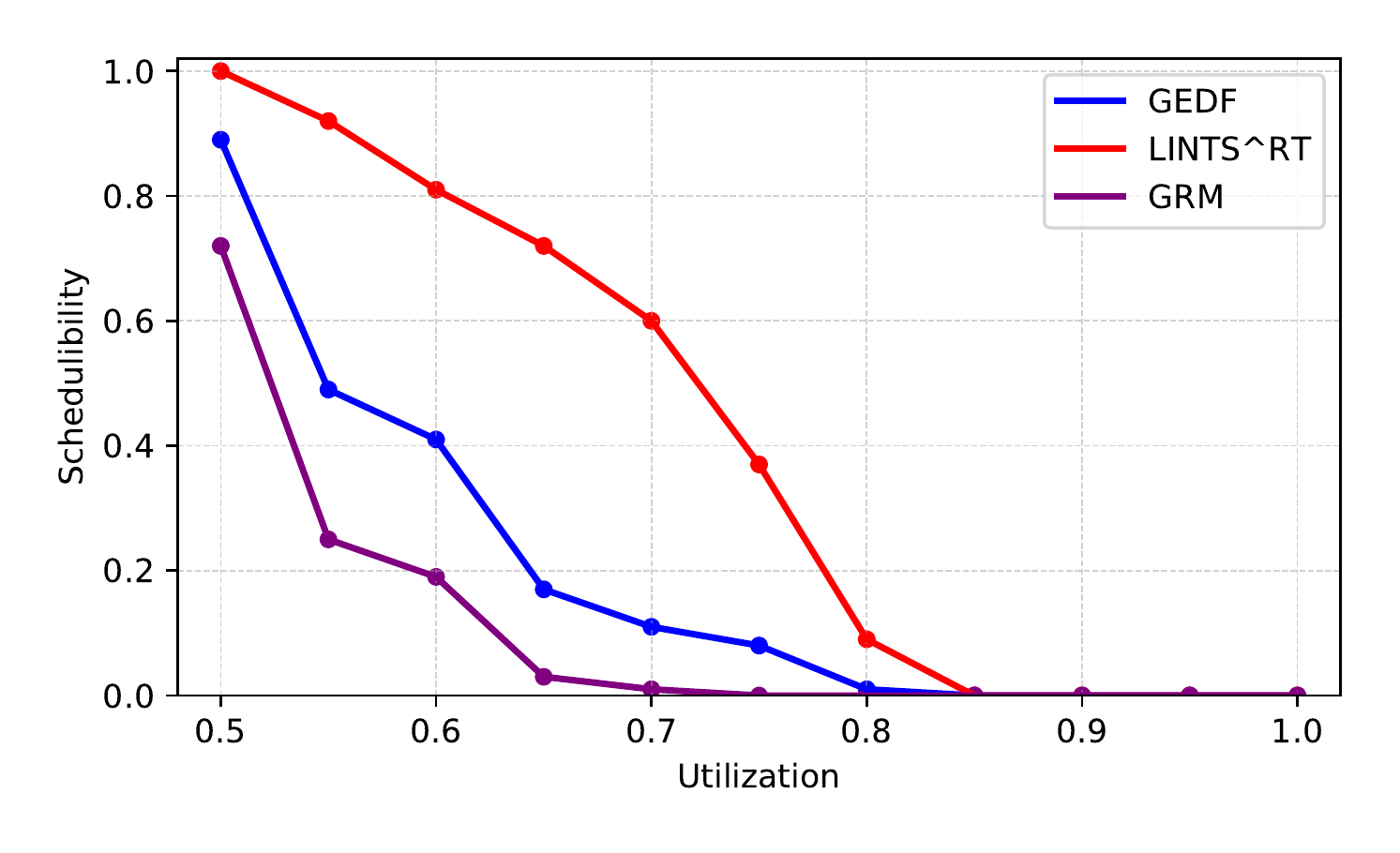}
        \caption{mixed.}
    \end{subfigure}

    \caption{Runtime schedulbility for ($m=4$, Non-preemptive, heterogeneous, \{light, medium, heavy, mixed\}) vs. $\frac{U}{m}$ }
    \label{fig:non-preemptive_heterogeneous}
\end{figure*}

\vspace{-2mm}
\section{Implementation and Evaluation}

We have implemented \name{} in a simulator for online configuration and LITMUS$^{RT}$ for offline configuration.
%We have conducted extensive schedulability experiments to evaluate our \name. 

\vspace{-2mm}
\subsection{Implementation details}
\label{sec:imp_details}

\subsubsection{Simulator Implementation}
\label{sec:simulator}
The simulator is used for both evaluation and MCTS roll-outs, and is responsible for releasing jobs, simulating job execution and computing system state. The simulator maintains a clock value $c$. For each run of the simulator, the following steps happen:
\begin{itemize}
    \item A random integer $r$ is generated.
    \item For each task $\tau_i$ in the task set, if there is no job released by $\tau_i$ and $mod(c, p_i) == r$, then release a new job for $\tau_i$.
    \item Compute the system state and send it to the scheduler.
    \item Update the remaining execution time for every job that was selected by the scheduler to execute.
    \item Remove completed jobs from the queue.
    \item Increment $c$.
\end{itemize}

In the future, we plan to extend this simulator to handle other kinds of hardware constraints such as I/O and memory. For MCTS roll-outs, multiple simulations need to run at each node, using the fact that these simulations are independent of each other they can be run in parallel. Hence we develop a simulator that is implemented in CUDA~\cite{nvidia2011nvidia} to utilize the capabilities of a GPU. Additionally, we can see that extending to non-preemptive and heterogeneous systems is intuitive.

% \subsubsection{MCTS Optimization}
% As we discussed in \ref{sec:mcts}, we need to run multiple MCTS roll-outs ($N_s$).  However, for a task set containing 10 tasks, an MCTS with $N_s = 20$ would need to simulate for $2 \times 10^{11}$ time slots on average. The simulator running on the CPU on the other hand, is only able to process $2.5 \times 10^5$ time slots per second per core, which means training the DNN with 100 task sets would at least take $10^6$ seconds, which is not acceptable. Since MCTS is a fully parallel algorithm, in order to improve the efficiency, we rewrite our simulator using CUDA~\cite{nvidia2011nvidia}, and run the MCTS simulation in parallel. After this optimization, we are able to process $8 \times 10^8$ time slots per second, which reduces the training duration to $10^4$ seconds (two orders of magnitude faster).

\vspace{-2mm}
\subsection{Evaluation and Training Setup}
\label{sec:evaluation_setup}
% The simulator can handle both preemptive and non-preemptive, as well as heterogeneous system simulations. Moreover, the simulator also takes several types of overheads into account. This enables us to explore diverse system configurations and parameter space.

We consider both preemptive and non-preemptive systems, and for each setting, we also consider platforms with $m \in \{2, 4\}$ processors of both homogeneous and heterogeneous architecture. 
For a given $m$, preemption setting and architecture setting, we vary the total utilization $U$ across $[0.5 \times m, m]$. We also separate the task sets into 4 different types based on their per-task-utilization: light, medium, heavy, and mixed. \say{Light} task sets contain tasks with utilization that varies from 0 to 0.2, \say{medium} task sets contain tasks with utilization from 0.2 to 0.5, for \say{heavy}, utilization is from 0.5 to 0.8, and \say{mixed} contains task with all the different utilization. 
Furthermore, for each configuration ($m$, preemption setting, architecture setting, total utilization $U$ and per-task utilization type), we randomly generate multiple task sets using the unbiased task set generator from \cite{emberson2010techniques}, which yields tasks whose utilization is uniformly distributed in the specified range for the configuration. Task periods were drawn uniformly at random from the set $\{50, 54, 60, 64, 72, 75, 80, 81, 90, 96, 100, 108, 120, 125, 128,\\ 135, 144, 150, 160, 162, 180, 192, 200\}$ (that only contains prime factors $\{2, 3, 5\}$), which reflects realistic timing constraints and ensures a short hyper-period. These generated task sets are used to train and evaluate \name. 
Also, note that the total utilization of our generated task sets is a little smaller than it is in the configuration (at most by 0.5 percent). 
For heterogeneous systems, we use the setting where half of the processors are working at a speed of \say{1}, and the remaining processors are working at a speed of \say{0.5}.
\begin{figure*}[t]
    \captionsetup[subfigure]{aboveskip=-4pt,belowskip=-3pt}
    \centering
    
    \begin{subfigure}[b!]{0.24\textwidth}
        \includegraphics[width=\textwidth]{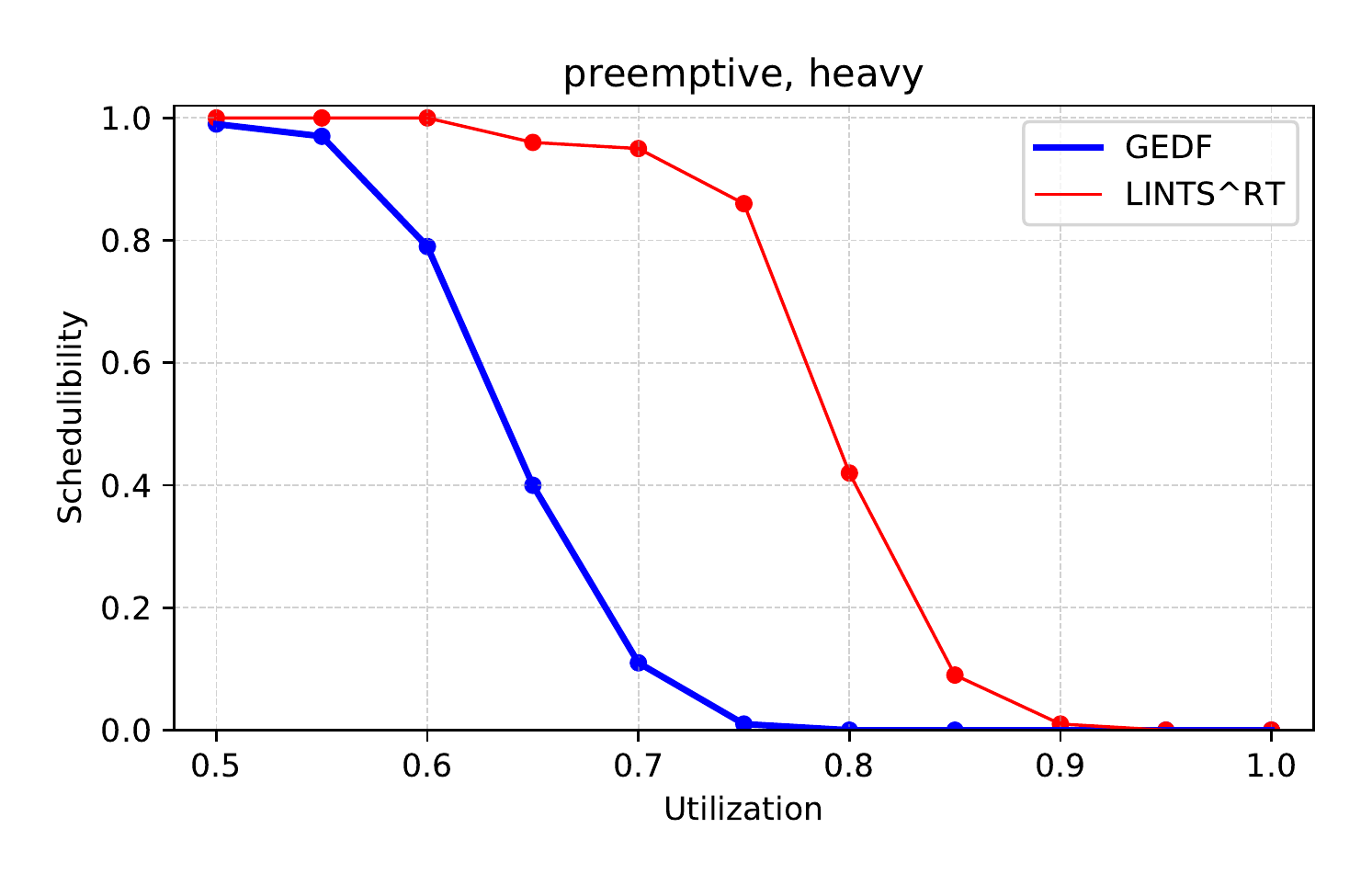}
        \caption{Heavy, preemptive.}
    \end{subfigure}
    \begin{subfigure}[b!]{0.24\textwidth}
        \includegraphics[width=\textwidth]{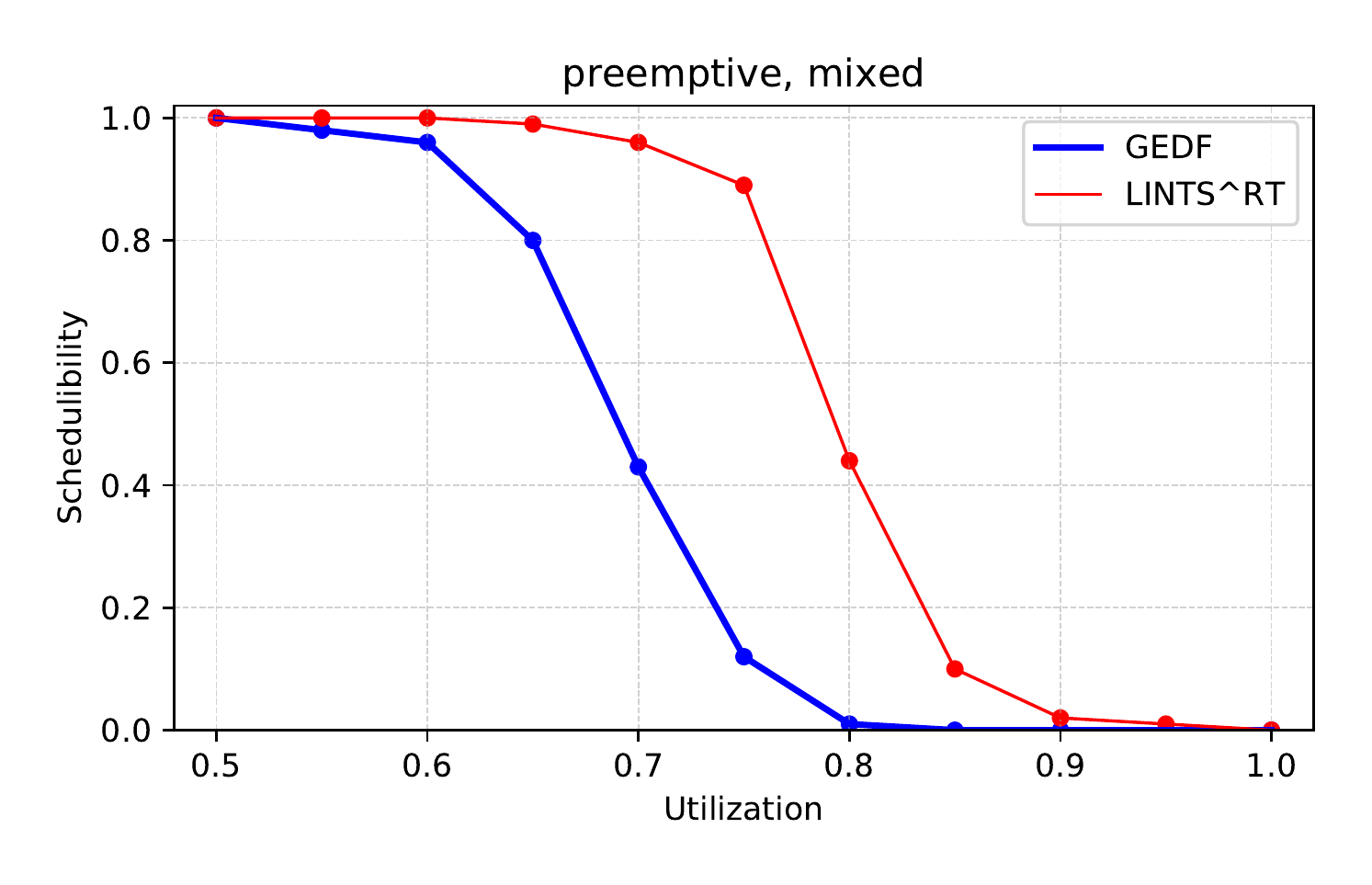}
        \caption{Mixed, preemptive.}
    \end{subfigure}
    \begin{subfigure}[b!]{0.24\textwidth}
        \includegraphics[width=\textwidth]{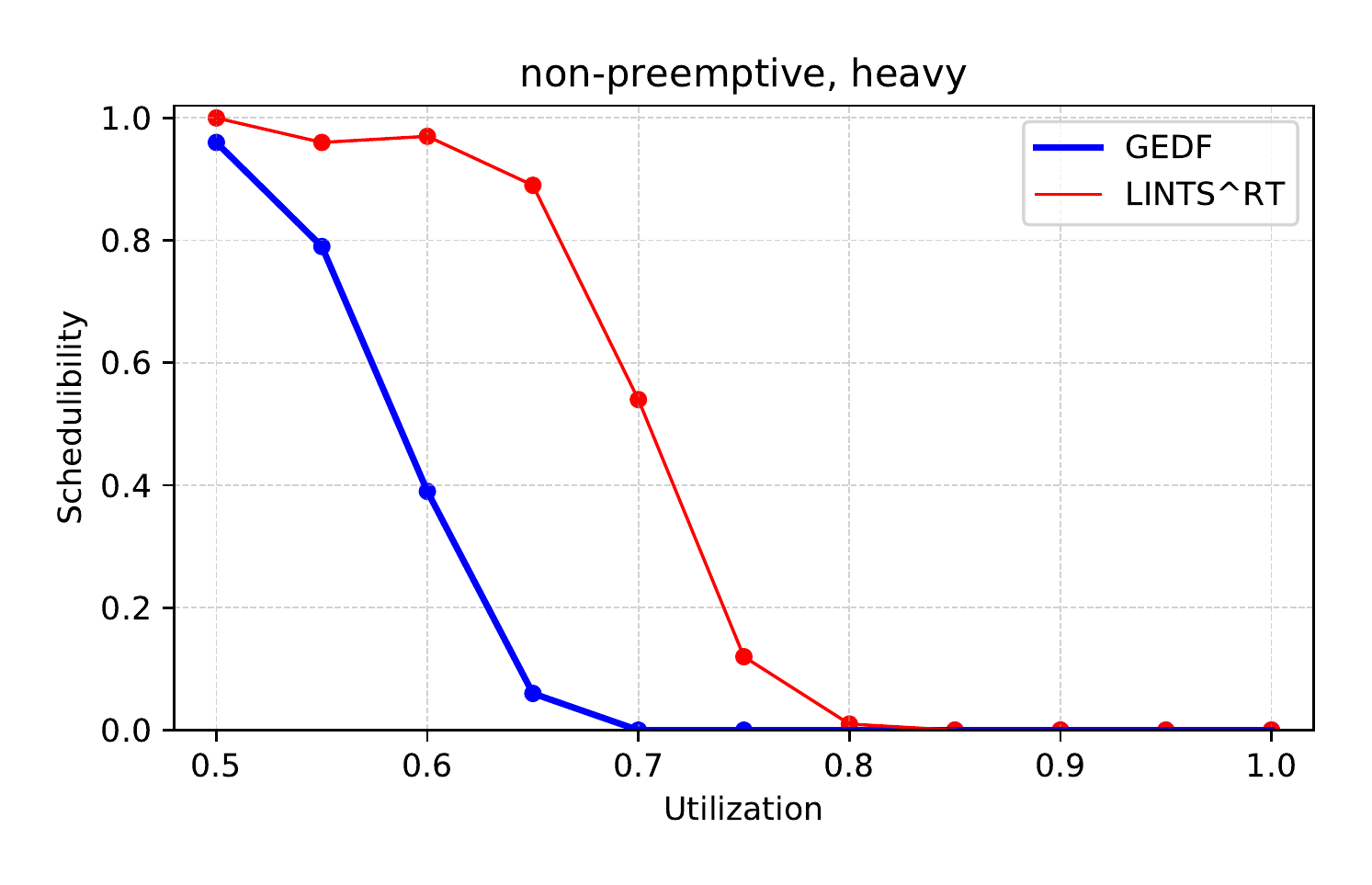}
        \caption{Heavy, non preemptive.}
    \end{subfigure}
    \begin{subfigure}[b!]{0.24\textwidth}
        \includegraphics[width=\textwidth]{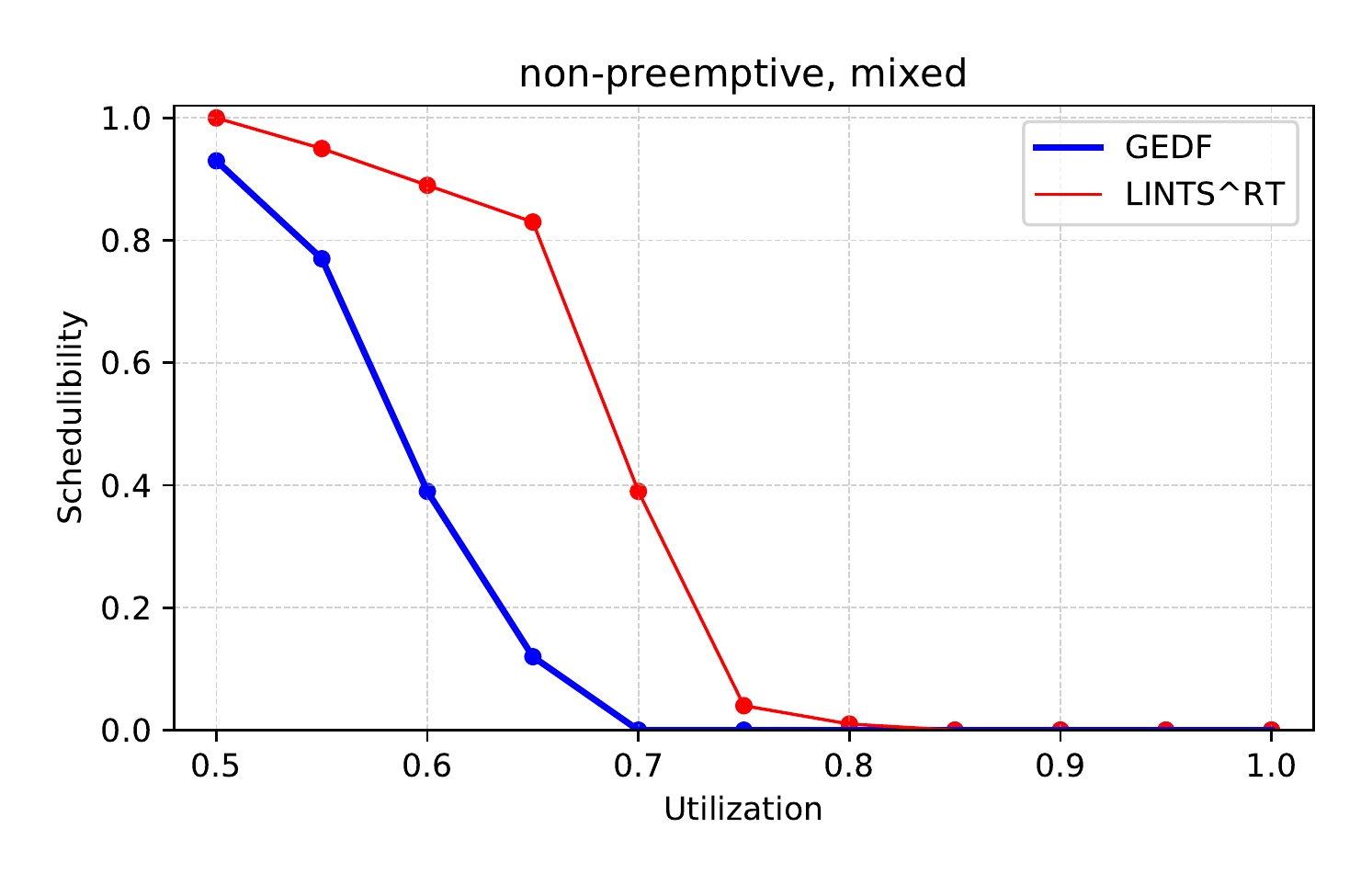}
        \caption{Mixed, non preemptive.}
    \end{subfigure}
    
    % \subfloat[Heavy, preemptive. ]{\includegraphics[width=0.24\textwidth]{figures/results/litmus_preemptive_heavy.pdf}}
    % \subfloat[Mixed, preemptive. ]{\includegraphics[width=0.24\textwidth]{figures/results/litmus_preemptive_mixed.pdf}}
    % \subfloat[Heavy, non-preemptive. ]{\includegraphics[width=0.24\textwidth]{figures/results/litmus_non-preemptive_heavy.pdf}}
    % \subfloat[Mixed, non-preemptive.]{\includegraphics[width=0.24\textwidth]{figures/results/litmus_non-preemptive_mixed.pdf}}
    \caption{Runtime schedulbility (LITMUS$^{RT}$) for  ($m=2$, \{preemptive, Non-preemptive\}, homogenous, \{heavy, mixed\}) vs. $\frac{U}{m}$}
    \label{fig:limus-rt}
       \vspace{-4mm}
\end{figure*}
\subsubsection{DNN Training}
% We implement our DNN on a mature machine learning framework PyTorch~\cite{paszke2017automatic}. On PyTorch, the DNN discussed in Sec.~\ref{sec:DNNTrain} takes around 10 nanoseconds to execute on a GPU (NVIDIA GTX 2080 Ti) and 4 milliseconds on a CPU with 8 cores. 

% We train our DNN using the approach described in Sec.~\ref{sec:DNNTrain} for a single system configuration ($m$, preemption setting, architecture setting, total utilization $U$ and per-task utilization) with 100 task sets. Since the system is a sporadic task system, the period and the execution times are subject to a uniform probability distribution.
We implement our DNN on a mature machine learning framework PyTorch~\cite{paszke2017automatic}. We train our DNN using the approach described in Sec.~\ref{sec:DNNTrain} for each system configuration, which is a tuple of ($m$, preemption setting, architecture setting, per-task-utilization type). Training is done on 100 task sets generated using the methods discussed in the previous section. Sporadicity of the task model is handled by the simulator, as discussed in section \ref{sec:simulator}.

The machine used to train the deep neural network contains an \textit{Intel(R) Xeon(R) CPU E5-2650 v4 @ 2.20GHz} and two NVIDIA GTX 2080 Ti GPUs. 
DNN parameters are updated using the stochastic gradient descent method (SGD) with momentum and learning rate annealing. The loss function is calculated according to Equation~\ref{eq:loss}. Table~\ref{tab:lr} shows the learning rate of the policy and value networks over different training phases. Parameters for the networks are updated over 10000 mini-batches containing 256 samples each, which are selected uniformly from the search trees.
\begin{table}[!h]
    \centering
    \begin{tabular}{ccc}
        \hline
        Simulation & Learning Rate for Policy & Learning Rate for evaluation \\
        \hline\hline
        0-500 & 0.1 & 0.01 \\
        500-1000 & 0.01 & 0.001 \\
        1000-1500 & 0.001 & 0.0001 \\
        1500-2000 & 0.0001 & 0.00001 \\
        \hline
    \end{tabular}
    \caption{Learning rates.}
    \vspace{-2mm}
    \label{tab:lr}
\end{table} 

As discussed earlier, each pass through the DNN schedules one job, which takes around 10 nanoseconds to execute on a GPU (NVIDIA GTX 2080 Ti) and 4 milliseconds on a CPU with 8 cores. This makes the overhead of \name{} = $10 \times m$ (20ns, 40ns for $m=\{2, 4\}$ respectively).
% The system configuration for the training is as follows: $m = 4$, preemptive, homogeneity, and $N_s = 20$ simulations using MCTS, with training using the SGD. During the training, parameters are updated by over $10000$ mini-batches of samples of size 256, which are selected uniformly from the search trees. 

\vspace{-2mm}
\subsection{Results}
Since \name makes scheduling decisions based on the current system state, coming up with a static schedulability test is difficult. So, we evaluate \name based on the runtime schedulability, i.e., the percentage of 1000 generated task sets that are schedulable by the scheduler for each system configuration ($m$, preemption setting, architecture setting, $U$, per-task-utilization distribution). Following results show graphs with the y-axis representing the runtime schedulability of each scheduler and the x-axis representing the total utilization of task set divided by the number of processors, i.e., $\frac{U}{m}$.

\subsubsection{Results on Preemptive, Homogeneous / Heterogeneous Settings}
Fig.~\ref{fig:preemptive_homogeneous} shows the results of \name{} working on a preemptive and homogeneous system with processor count $m= 4$.
As seen in the figure, \name{} yields a much higher runtime schedulability compared to GEDF and GRM, particularly when per-task utilizations are heavy. We performed experiments for $m=\{2,4\}$. Due to space constraints, we only show the results for the 4-processor case. However, the general trend for both 2-processor and 4-processor cases is similar, with all the schedulers performing slightly better in the 2-processor case.
% Another interesting observation is that \name{} yields a better performance on the 2-processor system compared to the 4-processor case. 
The reason for this in case of \name is that, at the beginning of every time slot, the DNN needs to be executed for each of the processors. Thus, the latency overhead due to the DNN decision-making process will increase along with an increasing number of processors. 
Under heterogeneous settings, As seen in Fig.~\ref{fig:preemptive_heterogeneous_8}, \name{} still outperforms GEDF and GRM by a large margin.

\subsubsection{Results on Non-preemptive, Homogeneous / Heterogeneous Settings}
For non-preemptive settings, Figs.~\ref{fig:non-preemptive_homogeneous} and \ref{fig:non-preemptive_heterogeneous} show that \name{} can outperfom GEDF and GRM by a significant margin. 
The reason is that MCTS is a search-based method, which takes trial-and-error actions while finding the feasible schedules, and the neural network is trained from the data generated by MCTS. Therefore, the neural network can learn  patterns of job selections that are able to reduce the number of missed deadlines. This result is encouraging because non-preemptive scheduling on either homogeneous or heterogeneous multiprocessors is known to be a notoriously hard problem.

It is clear from the results that the acceptance rate of the \name is much higher than GEDF and GRM. Additionally, empirical observations show that task sets accepted by either GEDF or GRM are also accepted by \name. From this, we can conclude that \name offers significant optimizations with no detrimental side-effects over the traditional scheduling approaches. Immediate future work would be formulating a utilization test using the weights of the DNN as parameters in its schedulability test since a trained DNN is a deterministic component which, when given the same system state, would always provide a consistent scheduling decision.

\vspace{-2mm}
\subsection{Experiments in LITMUS$^{RT}$}

We also perform evaluation on LITMUS$^{RT}$ through table-driven scheduling tools of LITMUS$^{RT}$ for periodic task sets.\footnote{We note that supporting sporadic task scheduling using \name{} within LITMUS$^{RT}$ requires to integrate the entire machine learning framework in the OS kernel, which consists of a large number of mathematical libraries, such as tensor calculation libraries and automatic differentiation libraries. As this effort would be significant, we leave this implementation as an immediate future work.} Due to space constraints, we omit some discussion of implementation details. 
Table-driven scheduling is a part of LITMUS$^{RT}$, and it is realized as a reservation type in the \textit{P-RES} plugin. 
On multiprocessor systems, table-driven reservations are partitioned, which means each reservation is restricted to a scheduling slot on only one processor. Thus, we use different table-driven reservations on each core. 
For the sake of comparison, we still choose to compare against GEDF, which is implemented using the \textit{GSN-EDF} plugin. 

As discussed earlier, the simulator is important for \name{} to perform well. If the simulator does not reflect the real environment of a platform, both the trained neural network and the generated static schedule will not work well. 
Our way to handle overheads is to add an empirical value $\bar{o}$ to the original execution time. We would like to use this evaluation to answer two questions: Is our estimation of the total overhead effective? Will the generated static schedule outperform the GEDF-based scheduler implemented within LITMUS$^{RT}$ Fig.~\ref{fig:limus-rt} shows the experimental results. As we can see in Fig.~\ref{fig:limus-rt}, \name{} outperforms GEDF by a wide margin for both preemptive and non-preemptive settings, which means that considering the empirically-measured overheads in the training phase works well and facilitates \name{} to provide overhead-conscious runtime scheduling decision making.

\section{Related Works}
As we have discussed the extensive set of related works on real-time scheduling in Sec.~\ref{sec:background}, we focus herein on discussing works related to applying machine learning techniques in application domains relevant to resource management. DeepRM  \cite{mao2016resource} is of particular relevance.  DeepRM applies RL techniques to solve the problem of resource management in online scheduling. It was proposed as an online resource management scheduler for compute clusters, cloud computing, and video streaming applications. Even though DeepRM is an online scheduler, the latency constraints for its application domain are very loose, which makes the design of DeepRM totally different from \name{}, which expects to make real-time scheduling decisions. Additionally, their approach doesn't consider any non-determinism, such as sporadic task releases, while calculating the reward for their actions. As discussed in Sec. \ref{sec:design}, addressing these differences is non-trivial. We are not aware of any work that applies reinforcement learning for dynamic-priority real-time scheduling.

\section{CONCLUSION}

In this paper, we present \name{}, a learning-based testbed for intelligent real-time scheduling, which has the potential to handle various workload and hardware complexities that are hard to handle in practice. We first present a basic design of \name{} for supporting sporadic workloads on homogeneous multiprocessors, and then demonstrate how to easily extend the framework to handle further complexities in the form of non-preemptivity and resource heterogeneity. 
Both application- and OS-level implementation and evaluation demonstrate that \name{} is able to achieve significantly higher runtime schedulability under different settings compared to perhaps the most commonly applied schedulers, global EDF and RM. %To our knowledge, this work is the first attempt to design and implement an extensible learning-based testbed for autonomously making real-time scheduling decisions. 

%In this paper, we present \name{}, a learning-based testbed for intelligent real-time scheduling which has the potential to handle various workload and hardware complexities seen in practice. We first present a basic design of \name{} for supporting sporadic workloads on homogeneous multiprocessors, and then demonstrate how to easily extend the framework to handle further complexities in the form of non-preemptivity and resource heterogeneity. Both application- and OS-level implementation and evaluation demonstrate that \name{} is able to achieve significantly higher runtime schedulability under different settings compared to perhaps the most commonly applied schedulers, global EDF and RM. To our knowledge, this work is the first attempt to design and implement an extensible learning-based testbed for autonomously making real-time scheduling decisions. 

\bibliographystyle{IEEEtran}
\bibliography{main}

\end{document}